\newif\ifOneCol

\ifOneCol
\documentclass[draftcls,12pt,onecolumn]{IEEEtran}
\else
\documentclass[journal]{IEEEtran}
\fi
\IEEEoverridecommandlockouts

\usepackage[noadjust]{cite}
\usepackage{amsmath,amssymb,amsfonts}
\usepackage{algorithmic}
\usepackage{graphicx,color}
\usepackage{textcomp}
\usepackage{mathtools}
\usepackage{array}
\usepackage{bm}
\usepackage{algorithm}
\usepackage{mathrsfs}
\usepackage{enumerate}
\usepackage{ragged2e}
\usepackage{filecontents}
\usepackage{url}
\usepackage{times}
\usepackage{multirow}
\usepackage{subfigure}
\usepackage[caption=false]{subfig}

\def\cell{{\text{cell}}}
\def\per{{\text{per}}}
\def\step{{\text{step}}}
\def\rsrp{{\text{RSRP}}}
\def\posi{{\text{pos}}}
\def\nega{{\text{neg}}}
\def\itapx{{\texttt{sTAV}}}
\def\itaps{{\texttt{sTAP}}}

\ifCLASSINFOpdf
\else
\fi

\begin{document}

\title{Smart Timing Synchronization for \\ Small Data Transmission}

\author{Gautham~Prasad,
        Nadhem~Rojbi,
        Flynn Dowey,
        Nikhileswar Kota,
        Lutz Lampe,~\IEEEmembership{Senior Member,~IEEE},
        and Gus Vos,~\IEEEmembership{Senior Member,~IEEE}%
\thanks{Gautham Prasad is with Ofinno, Reston, VA, USA. Nadhem Rojbi was with the Department of Electrical and Computer Engineering, Vancouver, BC, Canada. Lutz Lampe is with the Department of Electrical and Computer Engineering, The University of British Columbia, Vancouver, BC, Canada. Gus Vos is with Sierra Wireless, Richmond, BC, Canada. Email: gautham.prasad@alumni.ubc.ca, rojbi.nadhem@gmail.com, lampe@ece.ubc.ca. Gus Vos is with Sierra Wireless Inc., Richmond, BC, Canada. Email: gvos@sierrawireless.com}
\thanks{This work was supported in parts by the Natural Sciences and Engineering Research Council of Canada (NSERC) and Sierra Wireless Inc., Richmond, BC, Canada. A part of this work has been submitted to the United States Patent and Trademark Office for a utility patent~\cite{prasad2022method}.}}

\markboth{}{}

\maketitle


\begin{abstract}
Cellular Internet-of-things (C-IoT) user equipments (UEs) typically transmit periodic but small amounts of uplink data to the base station. \textcolor{black}{To avoid undergoing a traditional random access procedure prior to every transmission, 5th generation (5G) and newer systems use configured grants for small data transmission (CG-SDT), which is equivalent to its long-term evolution (LTE) counterpart of preconfigured uplink resources (PURs)-based transmission. CG-SDT configures uplink resources to UEs in advance for transmission without a random access procedure. A prerequisite for CG-SDT is that the UEs must use a valid timing advance (TA). This is done by validating a previously held TA before CG-SDT. While this validation is trivial for stationary UEs, mobile UEs often encounter conditions where the previous TA is no longer valid and a new one is to be requested by falling back to legacy random access procedures. This limits the applicability of CG-SDT in mobile UEs. To this end, we propose UE-native smart timing synchronization techniques to counter this drawback and ensure a near-universal adoption of CG-SDT. We introduce new machine learning-aided solutions for validation and prediction of TA for UEs with any type of mobility. We perform comprehensive simulation evaluations across different types of communication environments to demonstrate the effectiveness of our proposed solution in predicting the TA.}
\end{abstract}

\begin{IEEEkeywords}
Internet of Things (IoT), machine type communication (MTC), configured grant small data transmission (CG-SDT), reference signal received power (RSRP), timing advance (TA). 
\end{IEEEkeywords}

\section{Introduction} \label{Introduction}
\IEEEPARstart{B}{illions} of devices that can collect and transmit data, including wearables, smart meters, intelligent vehicles, and next-generation transportation systems, to name a few, are being connected to the Internet~\cite{vaezi2022cellular}. With a large growth in such interconnected devices, i.e., the Internet-of-things (IoT), there is a constant need for innovation in Low-Power Wide-Area (LPWA) cellular technologies to ensure massive connectivity and enhanced energy efficiency for IoT devices~\cite{vaezi2022cellular, chettri2019comprehensive, akpakwu2017survey}. \textcolor{black}{The 3rd Generation Partnership Project (3GPP) has historically addressed the requirements of LPWA cellular networks in the long-term evolution (LTE) for machine type communication (LTE-M) and narrowband IoT (NB-IoT) specifications~\cite{zayas20173gpp, hoglund2017overview, hassan2020nb, althumali2018survey}, which have been regularly improved to meet the needs of growing IoT networks~\cite{ericsson_iot, gsma_5G}. In addition, the 5th generation (5G)/5G-Advanced systems have continued to support small data transmission (SDT) for LPWA applications~\cite{khlass2021efficient, xin2021small, tr_38331}.}

IoT user equipments (UEs) are typically battery powered and supplied as a single charge version~\cite{callebaut2021art, ratasuk2016nb, lu2017reaching}. Hence, to ensure longer battery life, cellular IoT (C-IoT) UEs generally remain idle, employing discontinuous reception (DRX), extended discontinuous reception (eDRX), and/or power saving mode (PSM) \cite{ras_power}. Such an operation is suitable for IoT applications, as IoT traffic typically consists of short and infrequent blocks of data. However, a C-IoT UE that is in an \textit{idle mode} must establish a connection and enter a \textit{connected active mode} to transmit data to the base station~\cite{ts_36331}. This connection establishment procedure consumes an extended period of time. Although this may be acceptable for mobile broadband applications that have extended lengths of data payload to transmit, this overhead is magnified in the case of C-IoT applications, which significantly reduces the battery life of these devices. 

Establishing connection in C-IoT networks involves a random access procedure, during which time, the network UEs compete with each other to obtain resources for bidirectional communication. A standard 4-step contention-based random access procedure involves at least four message exchanges between the UE and base station before a connection is established~\cite{althumali2018survey, li2015dynamic}. \textcolor{black}{To reduce this overhead, the use of random-access SDT (RA-SDT) was introduced to piggyback short uplink data together with random access messages~\cite{shi2023novel}. This procedure is also referred to as early data transmission (EDT) in LTE-based LTE-M and NB-IoT~\cite{hoglund20183gpp}. EDT or RA-SDT reduced network latency by up to three seconds and extended the battery life of C-IoT UEs by up to three years~\cite{hoglund20183gpp}. The gains are further increased with the use of configured grant small data transmission (CG-SDT), wherein the C-IoT UEs are configured resources (i.e., configured grants (CGs)) beforehand, which they can then use for uplink transmission of small amounts of data without undergoing any random access message exchange~\cite{khlass2021efficient, tr_38331, shi2023novel}. Its LTE-time counterpart is referred to as transmission on preconfigured uplink resources (PURs)~\cite{hoglund20203gpp}. One of the conditions for using the CGs for SDT is that the UE must apply a valid timing advance (TA) for its transmission, that is, the UE must advance its signal transmission by an amount dictated by the propagation delay between the UE and the base station. Failure to do so potentially causes interference between transmissions of UEs using adjacent resources and/or missed detection and the base station. Therefore, every UE must first validate its TA before attempting a CG-SDT. The TA held by the UE is obtained either during CG configuration or via a CG-SDT response following a successful CG-SDT. }

\subsection{Drawbacks with the State-of-the-art Designs}
TA validation is a trivial task for stationary UEs, since the propagation delay between the UE and the base station remains constant. However, supporting mobile C-IoT UEs requires designing TA validation strategies to minimize inter-UE interference. The use of mobile IoT devices in smart cities, vehicular communication links, and post-emergency networks~\cite{ayoub2018internet, surobhi2014iot, zanella2014internet}, among others, motivates the challenge of addressing TA validation for mobile devices in a C-IoT network. TA validation essentially involves determining whether a mobile UE has moved beyond a permissible movement (PM) that is dependent on the length of cyclic prefix used in orthogonal frequency division multiplexed (OFDM) transmission. \textcolor{black}{Current-day 3GPP specifications allow UEs to validate TAs using (a) a centralized approach using the CG-SDT time alignment timer configured centrally by the base station, and/or (b) a distributed approach where UEs validate the TA using the measured reference signal received power (RSRP)~\cite{tr_38321}.} The latter is a dynamic technique that does not require base station assistance and is motivated by the correlation between the measured RSRP and the UE-base station separation. As the UE moves away from the base station, the measured RSRP reduces due to increased path loss and the TA increases due to a larger propagation delay, and vice versa. Thus, the state-of-the-art RSRP-based TA validation method~\cite[Sec. 5.27]{tr_38321} uses a threshold-based approach, where the difference in the measured RSRP (in logarithmic scale) between the two locations of UE is compared with a positive and a negative threshold to determine whether the UE has moved beyond PM. 

We identify three issues with the state-of-the-art method. First, the technique is highly dependent on fixed thresholds that are set based on path loss models~\cite{rsrp_patent_gus, R4_1905499, R4_1905500}. This idealized assumption is not always reliable, and data-driven methods are preferable in such scenarios, as also identified in our prior work~\cite{rsrp_patent_gus}. We build on this idea and provide detailed investigation and methodologies for TA validation to expand the use of CG-SDT. Following our proposal in~\cite{rsrp_patent_gus}, data-driven methods for obtaining an indication of TA have also been investigated in the context of expediting random access~\cite{kim2021two}, along the lines of machine learning (ML)-based timing estimation proposed for various facets of communication systems~\cite{schmitz2019deep, ali20206g, wu2019deep}. \textcolor{black}{The maturity of ML-based timing and location estimation techniques has also advanced adoption of such methods in the 3GPP standards, with a new work-item on artificial intelligence (AI)/ML native air interface being specified in the upcoming Rel-$19$~\cite{lin2023bridge, 10467186, chen20235g}.}

A second issue with legacy threshold-based TA validation is that the computation of thresholds and error margins is optimized for several redundant initial UE positions. For example, when the UE is near cell-edge, further movement away from the base station sends the UE out of the cell. In such a case, the UE can no longer use the CGs\footnote{In this paper, references to CGs refer to CG-SDT grants, and not other configured grants that are intended, for example, for semi-persistent transmissions.} regardless of its TA. Recognizing such conditions is critical in improving error margins while designing threshold computations. 

\textcolor{black}{Finally, we identify a fundamental drawback with the idea of TA validation. When a UE determines a TA to be invalid, the legacy approach requires the UE to fall back to random access-based approaches like RA-SDT or legacy four- or two-step random access procedures to obtain a new TA. Therefore, mobile UEs that often move beyond the PM rarely perform CG-SDT. On the other hand, a C-IoT UE can always apply CG-SDT if, in place of validating the TA, it instead directly \textit{predicted} the TA at any given UE location. }

\textcolor{black}{In certain aspects, TA prediction is akin to location and uplink timing estimation, which have received considerable attention in the recent past (see, e.g.,~\cite{butt2021ml, lv2022deep, tian2024attention} and references therein for ML-based positioning and~\cite{fang2022deep, singh2024machinelearningbasedhybrid,jang2021deep} and references therein for timing estimation, among other representative works). However, TA prediction within the context of CG-SDT exhibits unique differences. First, location estimation is an unnecessarily complex manner of determining TA. For example, the value of TA for any UE in a fixed-radius-circle around the base station is the same, despite them being at different locations. As a result, while TA prediction requires knowledge of round-trip distance between a UE and base station, positioning requires estimation of absolute location of the UE regardless of where the base station is located. This is not only excessive, but as seen in~\cite{butt2021ml, lv2022deep, tian2024attention}, these techniques often substantially increase computational complexity of the inferring device. Second, TA prediction methods in the art, such as~\cite{fang2022deep, singh2024machinelearningbasedhybrid, jang2021deep} focus on TA prediction at the base station side by efficient detection of the physical random-access channel signal. Such a solution, while useful for initial access, is not applicable to CG-SDT scenarios, where the main goal of using CG-SDT is to avoid performing a random-access procedure. Furthermore, our focus is on a UE-side TA prediction solution, which operates without performing sequence detection but instead estimating the TA simply based on raw signal strength measurements such as reference signal received power values.  }

\subsection{Contributions}
\textcolor{black}{In light of the aforementioned constraints and} to counter the drawbacks we identified with the state-of-the-art TA validation~\cite[Sec. 5.27]{tr_38321}, we propose UE-native smart timing synchronization techniques. Our solution consists of two component choices, a smart TA validator (\itapx) and a smart TA predictor (\itaps). \itapx~is an enhanced TA validator that validates the previously held TA and extrapolates it as the current TA to be used for CG-SDT. When the latest TA held by the UE is deemed to be invalid, it mandates UEs to fallback on legacy TA acquisition methods as specified in~\cite{tr_38321}. \textcolor{black}{\itaps, on the other hand, is a UE-native implementation design that estimates the TA in an open-loop manner at any given UE location using its previous TA and the difference in RSRP, such that the UE can always use CG-SDT. A C-IoT UE could either use one of \itapx~or \itaps. Although the use of \itaps~is more attractive as it offers more than simply validating an existing TA, \itapx~may be applicable in scenarios where the network is incompatible with UEs proactively using an estimated TA. For instance, a 5G base station may not allow UEs to use any TA other than the one it configures the UE with so that inaccurate or rogue UEs do not intentionally or unintentionally cause network congestion due to the use of imprecise TAs.  } 

\textcolor{black}{Our proposed \itapx~is superior to the state-of-the-art TA validation technique~\cite[Sec. 5.27]{tr_38321} in two ways. First, our improved threshold determination strategy exploits the configuration nature of CG-SDT resources to reduce error margins. Second, our data-driven approach enhances validation sensitivity when compared to that achieved using the method in~\cite[Sec. 5.27]{tr_38321}. Further, our proposed \itaps~is different from similar ones found in the art (e.g., ~\cite{zhang2022machine, fang2022deep, singh2024machinelearningbasedhybrid,jang2021deep}) as described in Section I-A. As our results demonstrate in Section~\ref{sec:results}, we are able to achieve our target TA prediction rates with minimal impact to power consumption of C-IoT UEs. We list the contributions of our work in the following.}

\subsubsection{{Theoretical Contributions}}
\textcolor{black}{We propose an enhanced closed-loop and a closed-form expression based TA validation procedure that improves the performance of the state-of-the-art RSRP-based TA validation for CG-SDT. Our modified solution identifies when a UE is at the cell boundary or close to the base station and recognizes that there is no further movement away or toward the base station, respectively, that can invalidate the TA while remaining in the same cell. The new positive and negative validation thresholds that we develop using this prior knowledge enhances the detection rates and reduces false alarms in validating the TA.}

\textcolor{black}{We design an ML-aided solution for TA validation that requires less signaling overhead compared to the legacy \textit{threshold-based} approach. Furthermore, we advance the adoption of CG-SDT in 5G C-IoT networks by designing a TA prediction method. We first provide a closed-form expression for TA estimate computation based on the difference in RSRP and the previous TA. Additionally, we use supervised ML regression to design a predictor that can estimate the TA at any given location of the UE and for any movement that the UE has undergone. We then quantize the estimated floating-point TA for use during the CG-SDT. {Finally, we design sequential machine learning based solutions to use a multitude of historical TA values to predict the current propagation distance between the UE and the base station.}}

\subsubsection{\textcolor{black}{Evaluation Methodology}}
\textcolor{black}{We conduct a comprehensive evaluation campaign to investigate the performance of our proposed solutions. We perform 3GPP-compliant end-to-end link-level simulations under diverse communication environments and UE movements to evaluate the validation and prediction accuracy with using our proposed solutions. We demonstrate the robustness of \itapx~to operate in an environment-agnostic fashion and show that our method matches the detection rate and significantly reduces false alarms for TA validation when compared to the threshold-based approach. Similarly, we show that \itaps~is able to substantially reduce fallback rates using TA prediction in different types of channel environments. In addition, we present evaluation results to show the effectiveness of our results in translating to the real-world. We demonstrate this by evaluating \itaps~in an urban replica environment of Munich, Germany, with a base station mounted on top of its \textit{Frauenkirche} tower and a UE moving within about a one-mile radius around the base station. }

\subsection{Outline}
The rest of this paper is organized as follows. We describe the system model used in our work in Section~\ref{sec:system_model}. We propose our novel $\itapx$~and $\itaps$~solutions in Section~\ref{sec:itapx} and Section~\ref{sec:itaps}, respectively. In Section~\ref{sec:results}, we present the performance evaluation results of our proposed solutions. We reflect on our proposed methods to identify the application scenarios and potential implementation limitations of our methods in Section~\ref{sec:discussion}. We conclude our paper in Section~\ref{sec:conclusion}. Table~\ref{table:acronyms} lists acronyms and abbreviations used in this paper.

\textit{Notation}: Throughout the paper, the $\log$ operator represents logarithm to base $10$.

\begin{table}[!htp]
	\centering
	\caption{List of Acronyms and Abbreviations}\label{table:acronyms}
	\begin{tabular}{|c|l|}
		\hline
		Acronym/Abbreviation  		 	& Expansion \\ \hline \hline
		3GPP 				            & 3rd Generation Partnership Project \\ \hline
		5G 				                & 5th Generation \\ \hline
		AdaBoost  				        & Adaptive Boosting \\ \hline
        AI                              & Artificial Intelligence \\ \hline
		C-IoT 			                & Cellular Internet-of-Things \\ \hline
		CG		                        & Configured Grant \\ \hline
		CG-SDT 		                    & CG-Small Data Transmission \\ \hline
        CRS                             & Cell-specific Reference Signal \\ \hline
        DRX                             & Discontinuous Reception \\ \hline
        eDRX                            & Enhanced Discontinuous Reception \\ \hline
        FN                              & False Negative \\ \hline
        FNR                             & False Negative Rate \\ \hline
        FP                              & False Positive \\ \hline
        FPR                             & False Positive Rate \\ \hline
        IDRX                            & Idle Discontinuous Reception \\ \hline
        IoT                             & Internet-of-Things \\ \hline
        L2Boost                         & Least Squares Boosting \\ \hline
        LCM                             & Life Cycle Management \\ \hline
        LPWA                            & Low-Power Wide-Area \\ \hline
        LTE                             & Long Term Evolution \\ \hline
        LTE-M                           & LTE-Machine Type Communication\\ \hline
        ML                              & Machine Learning \\ \hline
        MTC                             & Machine Type Communication \\ \hline
        NB-IoT                          & Narrowband Internet-of-Things \\ \hline
        OFDM                            & Orthogonal Frequency \\
                                        & Division Multiplexing \\ \hline
        PM                              & Permissible Movement \\ \hline
        PSM                             & Power Saving Mode \\ \hline
        PSS                             & Primary Synchronization Signal \\ \hline
        PUR                             & Preconfigured Uplink Resource \\ \hline
        RA-SDT                          & Random Access \\
                                        & Small Data Transmission \\ \hline
        RE                              & Resource Element \\ \hline
        RMa                             & Rural Macro \\ \hline
        RSRP                            & Reference Signal Received Power \\ \hline
        SDT                             & Small Data Transmission \\ \hline
        SNR                             & Signal to Noise Ratio \\ \hline
        SSB                             & Synchronization Signal Block \\ \hline
        SSS                             & Secondary Synchronization Signal \\ \hline
        sTAP                            & Smart Timing Advance Predictor \\ \hline
        sTAV                            & Smart Timing Advance Validator \\ \hline
        SVM                             & Support Vector Machine \\ \hline
        TA                              & Timing Advance \\ \hline
        TDL                             & Tapped Delay Line \\ \hline
        TN                              & True Negative \\ \hline
        TNR                             & True Negative Rate \\ \hline
        TP                              & True Positive \\ \hline
        TPR                             & True Positive Rate \\ \hline
        UE                              & User Equipment \\ \hline
        UMa                             & Urban Macro \\ \hline
        UMi                             & Urban Micro \\ \hline
	\end{tabular}
\end{table}

\section{System Model} \label{sec:system_model}

\begin{figure}[t]
	\centering
	\includegraphics[width=8cm]{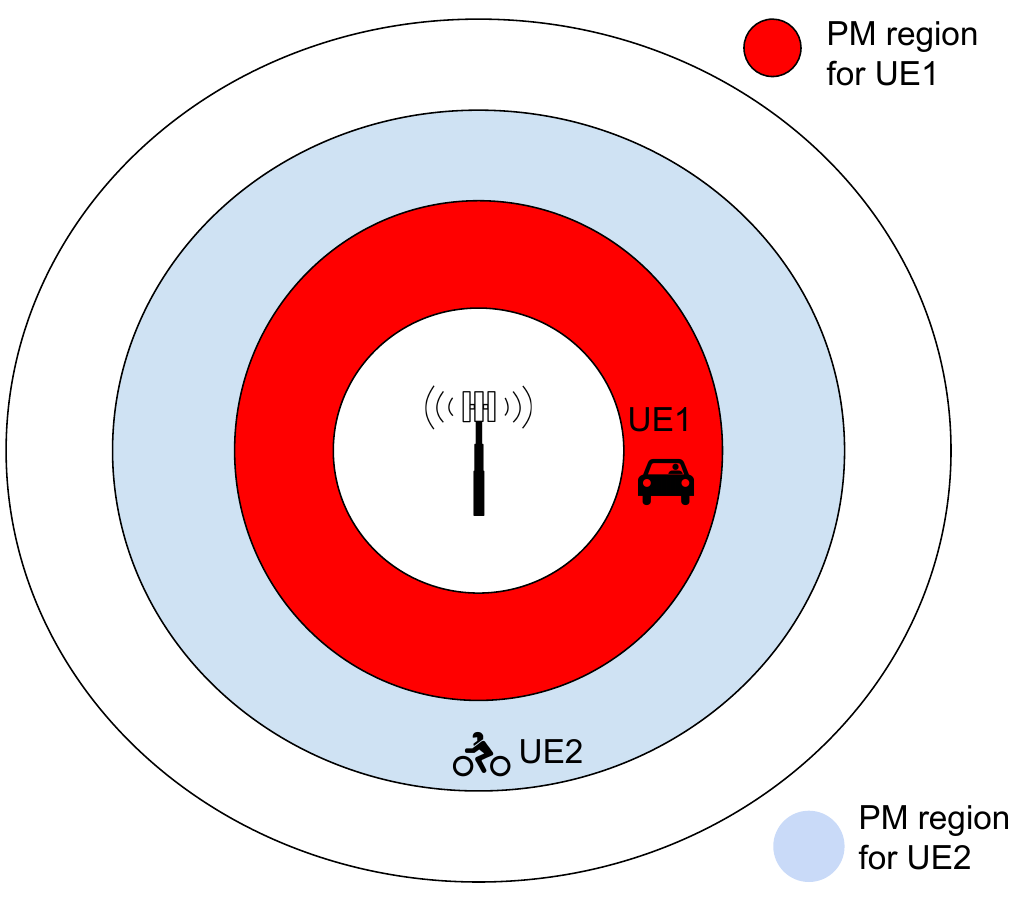} 
	\caption{An example illustration of two C-IoT UEs and their corresponding PM regions for using CG-SDTs with their previously held TAs. (Objects in the image are not to scale.)}
	\label{fig:SystemModel}
\end{figure}

We consider a C-IoT system, such as the one shown in Fig.~\ref{fig:SystemModel}, where a base station serves UEs within a cell. We model the cell as a Euclidean ball in $\mathbb{R}^2$, represented as 
\begin{equation} \nonumber
C_{r_\text{cell}} := \{x \in \mathbb{R}^2 : ||x||_2 \leq r_\text{cell}\},
\end{equation}
where $r_\text{cell}$ is the cell-radius. \textcolor{black}{Note that this is an approximation. Real-world cell coverage may not be circular~\cite{curry20205g, simic2017coverage}. Coverage varies depending on a variety of factors including antenna-type, physical environment around the base station, overlapping cell coverage, use of carrier aggregation, and other deployment factors, e.g.,~\cite{curry20205g, simic2017coverage, xu2020understanding, chiaraviglio2021massive, ye2024dissecting}.}

We consider $1.5~\text{km} \leq r_\text{cell} \leq 5~\text{km}$ to capture a range of possible deployment scenarios from an urban setting to a more rural environment. \textcolor{black}{Our choice of the lower end of cell size is based on TA validation for small cells (e.g., indoor distributed systems) being trivial due to the limited intra-cell mobility of UEs. Therefore, we do not consider such pico/femto cell scenarios or indoor environments. On the other hand, we choose the higher end of the cell size based on typical values of cell radius in macro-cells in real-world deployments, with no line-of-sight restrictions~\cite{maccartney2017rural}.}

C-IoT traffic is typically sporadic and uplink-heavy in nature, with data packet inter-arrival times commonly ranging between few seconds to a day~\cite[Annex. E]{tr_45820},~\cite[Annex. A]{tr_36888}. To save battery between transmissions, C-IoT UEs enter a \textit{sleep} state using DRX. Accordingly, the UE enters an idle DRX (IDRX) mode after the completion of the message exchange with the base station and a subsequent wait period. Upon request for data transfer either by the base station on IDRX pages (mobile terminated data) or by the application for uplink data (mobile originated data), the UE enters an active connected mode, which requires an extended duration of time to complete the random access procedure prior to entering the active connected mode~\cite{hoglund20183gpp}. When the UE has a small amount of data to transmit, the uplink data can be piggybacked on one of the random access messages for faster completion using RA-SDT to achieve a battery life extension of about $1.5$ times~\cite{s24175706}. By using CG-SDT, battery life of a C-IoT UE can be increased by up to $3.5$ times when compared to legacy random-access based approaches~\cite{s24175706}.

\begin{figure}[t]
	\centering
	\includegraphics[width=5.5cm]{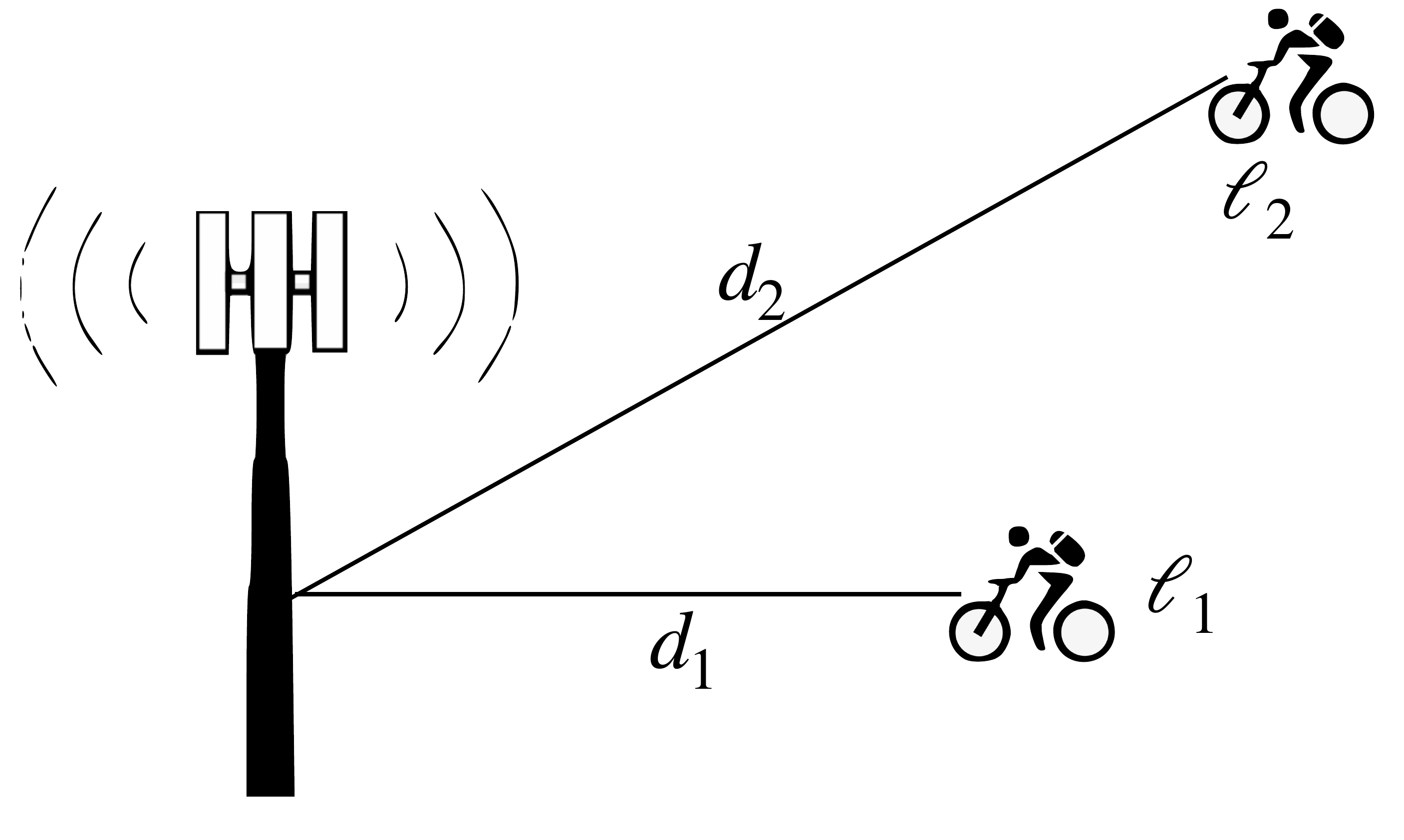} 
	\caption{A UE attempting PUR transmissions from two different locations at $\ell_1$ and $\ell_2$.}
	\label{fig:ue_positions}
\end{figure}

We consider a C-IoT network that uses CG-SDT, where the base station provides UEs with CG-SDT configurations as requested either by the network or the UEs themselves~\cite{hoglund20203gpp}. The UEs are allowed to use the CG-SDT configurations for CG-SDT so long as they remain in the same cell. Additionally, the CG-SDT configurations may also be configured with a CG-SDT time alignment timer, upon whose expiration, the TA held by the UE is no longer considered to be valid. Alternatively or additionally, the network may set the CG-SDT time alignment timer to a large value and allow the UEs to determine the validity of the TA independently by monitoring the RSRP. Static UEs, e.g., cellular-connectivity-enabled smart meters, can always use the TA that is provided during CG-SDT configuration. \textcolor{black}{However, TAs change with movement for mobile UEs such as terrestrial vehicles, mobile robots, aerial vehicles, and wearables~\cite{talavera2015mobile}.} As a result, the TAs must be validated before CG-SDT. We call the region within which the last available TA held by the UE is considered to be valid as the PM region, as shown in Fig.~\ref{fig:SystemModel}. As long as the UE is within this PM region when it attempts the next CG-SDT, it can continue to use the TA it received from the base station at the end of the previous CG-SDT. The width of this area is dependent on the extent to which an incorrect TA can be tolerated at the base station. This is in turn dependent on the length of the cyclic prefix used by the OFDM transmission. With a standard subcarrier spacing of $15$~kHz, the cyclic prefix spans about $4.7~\mu$s, which corresponds to a maximum permissible movement of $\Delta d_\per = 702$~m. \textcolor{black}{Notice that this PM does not dictate a general movement of the UE, but instead a movement that causes the UE-base station separation to increase more than $702$~m from its previous separation. For instance, if the UE moves along a circle with the base station as the center of the circle, irrespective of the actual distance traversed by the UE, the change in the UE-base station separation would be zero. }When the UE uses an invalid TA, i.e., when the UE uses its TA when its movement violates the condition, 
\begin{equation}\label{eq:movement_condition}
	\Delta d \leq \Delta d_\per,
\end{equation}
\textcolor{black}{where $\Delta d$ indicates the difference in the UE-base station distances between the current and the previous UE locations, its CG-SDT may interfere with UEs using adjacent resources. Consider an example of UE movement as shown in Fig.~\ref{fig:ue_positions}. At two different time instances (for example, at two CG-SDT instances), the UE is located at $\ell_1$ and $\ell_2$ that are at distances of $d_1$ and $d_2$ from the base station\footnote{Note that $d_1$ and $d_2$ represent straight-line distances between the UE and base station when the UE is at $\ell_1$ and $\ell_2$, respectively. This is equal to the distance traversed by the signal in an LOS environment. In an NLOS environment, the distances indicate the path lengths traversed by the signal and may be approximated as the UE-base station separations.}, respectively. In this scenario, }
\begin{equation}
    \Delta d = |d_2 - d_1|.
\end{equation}  

TAs are provided and updated by the base station in quantized amounts. The quantized version of the TA, $\tau_\text{q}$, can be represented as 
\begin{equation}\label{eq:ta_quant}
	\tau_\text{q} = \left \lfloor \frac{2d}{c \tau_\step} \right \rfloor,
\end{equation} 
where $d$ is the distance between the UE and the base station, $\tau_\step$ is the quantization step, which is equal to $0.52~\mu$s for a subcarrier spacing of $15$~kHz, and $c$ is the speed of light. Accordingly, we represent the deviation in $\tau_\text{q}$ corresponding to $\Delta d_\per$ as $\Delta \tau_{\text{q}, \per}$.

\section{TA Validation}\label{sec:itapx}

Legacy TA validation can be performed in two ways. The first is a centralized approach, where the base station provides the UE with a CG-SDT time alignment timer. The UE is then allowed to use CG-SDT resources, i.e., determine that the previously held TA by the UE is valid, when the CG-SDT-time alignment timer is running~\cite{tr_38321}. When the CG-SDT-time alignment timer expires, the UE no longer considers the TA to be valid. The base station determines the value of the CG-SDT-time alignment timer based the UE mobility and the periodicity of the CGs. A primary concern with such a centralized approach is that the movement of the UE is not always deterministic or uniform. Therefore, the base station often conservatively provides a lower value of CG-SDT-time alignment timer to ensure that the UE does not use an invalid TA. A better alternative to this method is a distributed approach, which allows the UEs to determine the TA validity based on the correlation between the TA and a measured power metric, such as RSRP. In the following, we focus on such an RSRP-based TA validation to improve the accuracy of detecting the TA validity.

\textcolor{black}{RSRP is computed as the linear average of the power contributions of resource elements (REs) that carry downlink reference signals (e.g., cell-specific reference signal (CRS) for LTE-based PUR implementations, and synchronization signal block (SSB) for 5G-based CG-SDT), measured over a certain bandwidth. The measurement bandwidth and the number of REs used to determine RSRP, however, is left to UE implementation~\cite{ts_36214}. The locations of the REs carrying the downlink reference signals (i.e., CRSs or SSBs) are determined using the physical ID of the camped cell, which is obtained when the UE is synchronized using the primary synchronization signal (PSS) and secondary synchronization signal (SSS)~\cite{ts_36211, tr_38331}. } The measured RSRP, $P_\rsrp$, can be represented as 
\begin{equation}\label{eq:rsrp_1st}
	P_\rsrp = \frac{P_\text{TX}}{\alpha(d_i)},
\end{equation}
where $P_\text{TX}$ is the downlink reference signal transmit power used by the base station and $\alpha(d_i)$ is the net power loss for a mobile UE that is located at $\ell_i$, which is at a distance $d_i$ from the base station. Across different types of environments,~\eqref{eq:rsrp_1st} can be represented in its general form as~\cite{tr_36814}
\begin{equation}
	P_{\rsrp,i} = \frac{P_\text{TX}}{\beta_i d_i^{k_i}},
\end{equation}
where $k_i$ is the path-loss exponent and $\beta_i$ is the scaling coefficient accounting for system losses at the $i$th location of the UE. Consider a UE movement from an initial location $\ell_{i-1}$ to a new one at $\ell_i$, as shown in Fig.~\ref{fig:ue_positions} (for $i=2$). The difference in RSRP in the logarithmic scale is
\begin{align}\label{eq:delRSRP_full}
	\Delta P_{\rsrp, i, \text{dB}} = \log \left( \frac{\beta_i d_i^{k_i}}{\beta_{i-1} d_{i-1}^{k_{i-1}}} \right).
\end{align} 
Assuming that the path-loss characteristics are the same at the two UE locations\footnote{\textcolor{black}{Recall that CG-SDT resources are valid only within the cell where the UE receives the CG-SDT configuration. As a result, $\ell_i$ and $\ell_{i-1}$ are within the same cell and therefore, likely to experience the same or substantially similar path-loss characteristics.}}, i.e., with $\beta_{i-1} = \beta_i$ and $k_{i-1}=k_i=k$,~\eqref{eq:delRSRP_full} can be simplified as 
\begin{equation}\label{eq:delRSRP_simple}
	\Delta P_{\rsrp, i, \text{dB}} = k \log \left( \frac{d_i}{d_{i-1}} \right).
\end{equation}
Since $P_\rsrp$ does not vary linearly with $d$, the observed $\Delta P_{\rsrp, \text{dB}}$ is not the same for a UE movement toward and away from the base station. Therefore, for a UE at the $i$th location, the threshold-based TA validation compares the measured $\Delta P_{\rsrp, \text{dB}}$ against a positive and a negative threshold, $\Delta P_{\max, \text{pos}}$ and $\Delta P_{\max, \text{neg}}$, respectively, which are computed as~\cite{rsrp_patent_gus}
\begin{equation} \label{eq:ThBasic}
 	\begin{split}
 		\Delta P_{\max, i, \text{pos}} &= k \log \left( 1+\frac{\Delta d_\per}{d_{i-1}} \right) - \epsilon_{\text{pos}} \\
 		\Delta P_{\max, i, \text{neg}} &= k \log \left( 1-\frac{\Delta d_\per}{d_{i-1}} \right) + \epsilon_{\text{neg}},
 	\end{split}
\end{equation}
where $\epsilon_{\text{pos}}$ and $\epsilon_{\text{neg}}$ are error margins used to account for measurement errors and channel variations. The error margins can be dynamically adjusted by the base station to achieve a desired level of validation accuracy, e.g., by sacrificing false alarms to improve detection rates.   

\subsection{Threshold-based \itapx}\label{subsec:threshold_itapx}
We incorporate~\eqref{eq:delRSRP_simple} and~\eqref{eq:ThBasic} as the baseline for building the threshold-based \itapx. Since $\Delta P_{\max, i, \text{pos}}$ and $\Delta P_{\max, i, \text{neg}}$ must always be positive and negative, respectively, and RSRP monotonically decreases with distance, the limits on $\epsilon_{\text{pos}}$ and $\epsilon_{\text{neg}}$ can be computed at the cell boundary as 
\begin{equation} \label{eq:ThMargin}
	\begin{split}
		0 &< \epsilon_{\posi} < k \log \left(1+ \frac{\Delta d_\per}{r_\cell}  \right) \\
		0 &< \epsilon_{\nega} < -k \log \left(1-\frac{\Delta d_\per}{r_\cell}   \right).
	\end{split}
\end{equation}
However, when the UE is close to the cell-edge, i.e., when $r_\cell - \Delta d_\per < d_{i-1} \leq r_\cell$, there is no movement away from the base station that will render the TA invalid. The UE either moves such that $\Delta d \leq \Delta d_\per$, in which case its TA is still valid, or moves such that $\Delta d > \Delta d_\per$ and exits the cell and can thus no longer use the CGs. Similarly, when the UE is close to the base station, i.e., when $d_{i-1} \leq \Delta d_\per$, there is no movement toward the base station that will render the TA to be invalid. Exploiting these two factors, \itapx~using threshold-based TA validation uses a modified detection approach as outlined in Algorithm~\ref{alg:itapx}. This also allows us to use larger error margins which are bounded by
\begin{equation}\label{eq:ThMargin_new}
	\begin{split} 
		0 &< \epsilon_{\posi} < -k \log \left(1-\frac{\Delta d_\per}{r_\cell}   \right) \\
		0 &< \epsilon_{\nega} < -k \log \left(1-\frac{\Delta d_\per}{r_\cell - \Delta d_\per}   \right).
	\end{split}
\end{equation}
Note that for Algorithm~\ref{alg:itapx}, \itapx~uses $d_{i-1}$ that is obtained using~\eqref{eq:ta_quant} with $\tau_{\text{q},i-1}$, where $\tau_{\text{q},i-1}$ is the actual TA value provided by the base station when the UE completes a successful CG-SDT while being located at $l_{i-1}$.
We show in Section~\ref{sec:results} that \itapx~in Algorithm~\ref{alg:itapx} with~\eqref{eq:ThMargin_new} matches the performance of legacy TA validation to detect invalid TA conditions and significantly improves the performance of identifying valid ones.
 
\begin{algorithm}[t]
	\centering
	\caption{The proposed threshold-based \itapx.} \label{alg:itapx}
	\begin{algorithmic}
		\STATE Compute $\Delta P_{\max, i, \text{pos}}$ and $\Delta P_{\max, i, \text{neg}}$ using~\eqref{eq:ThBasic}
		\STATE \textbf{while} CG-SDT data to transmit  \textbf{do}:
		\STATE Detect CRS and measure $\Delta P_{\rsrp, i, \text{dB}}$ 
		\IF{$d_{i-1} < \Delta d_{\per}$}
		\STATE	TA valid \textbf{if}: $\Delta P_{\rsrp, i, \text{dB}} < \Delta P_{\max, i, \text{pos}} $ 
		\ELSE 
		\IF{$d_{i-1} > r_{\cell} - \Delta d_\per$}
		\STATE TA valid \textbf{if}: $\Delta P_{\rsrp, i, \text{dB}} > \Delta P_{\max, i, \text{neg}}$ 
		\ELSE
		\STATE TA valid \textbf{if}: $\Delta P_{\max, i, \text{neg}} \! < \! \Delta P_{\rsrp, i, \text{dB}} \! < \! \Delta P_{\max, i, \text{pos}}$ 
		\ENDIF
		\ENDIF
	\end{algorithmic}
\end{algorithm}
 
\subsection{ML-aided \itapx}
Although the threshold-based \itapx~improves the performance of TA validation when compared to the state-of-the-art~\cite[Sec. 5.27]{tr_38321}, it still suffers from measurement inaccuracies due to the impact of noise at low signal-to-noise ratio (SNR) conditions, fading effects of the channel, and deviation of the practical environment behavior from the path-loss model used in computing the thresholds. To address this model deficit, we explore data-driven approaches using ML for TA validation. Our goal is to develop a robust \itapx~scheme, where a machine trained with synthetic data generated using path-loss models can operate in any typical cellular channel and noise environment. Training the machine with synthetic data gives us the freedom to use as many data samples as required for training to achieve the desired performance. 
 
\subsubsection{Formulating the Problem}
We formulate the problem as a supervised binary classification task, where a classifier determines if the previously held TA is valid or not based on the difference in RSRP, i.e.,
\begin{align}\nonumber
	  \hat{\tau}_{\text{q}, i} &=  \itapx(\tau_{\text{q}, i-1}, \Delta P_{\rsrp, i, \text{dB}})\\\label{eq:itapx_output} &=
	  \begin{cases}
	   \tau_{\text{q}, i-1}, &\text{if valid} \\
	   \infty, &\text{if invalid},
	  \end{cases}
\end{align}
where $\hat{\tau}_{\text{q}, i}$ is the output TA of \itapx~at the $i$th CG-SDT instance.
A result of $\hat{\tau}_{\text{q}, i} = \infty$ indicates to the UE that the previous TA is invalid. In such cases, the UE must fall back to other downward compatible network access methods, e.g., RA-SDT, for data transmission and TA acquisition.

We characterize the performance of \itapx~in terms of the true negative rate (TNR), $p_\text{TN}$, and the true positive rate (TPR), $p_\text{TP}$, of the classifier. We define true negatives as the cases where \itapx~indicates that $\hat{\tau}_{\text{q},i} = \infty$, when the TA is actually invalid. Therefore, the false positive rate (FPR), $p_\text{FP}$ can be written as 
\begin{equation}\label{eq:FPR}
    p_\text{FP} = 1 - p_\text{TN},
\end{equation}
indicating the rate of false positives where the TA is determined to be valid by \itapx~when the TA is in fact invalid. Along the same lines, we define true positives as conditions where \itapx~indicates the previously held TA to be valid when the TA is valid, and consequently, the false negative rate (FNR), $p_\text{FN}$, can be defined as 
\begin{equation}\label{eq:FNR}
    p_\text{FN} = 1 - p_\text{TP}.
\end{equation}
We summarize the conditions and outcomes of \itapx~in Table~\ref{table:classifier_outcomes}.
\begin{table}[t]
	\centering
	\caption{Conditions and Outcomes of {\itapx}}\label{table:classifier_outcomes}
	\begin{tabular}{|l|l|l|}
		\hline
		Condition 			& Actual TA 	& Predicted TA 	\\ \hline \hline
		True Positive (TP) 	& Valid 		& Valid 		\\ \hline
		True Negative (TN)	& Invalid 		& Invalid 		\\ \hline
		False Positive (FP) & Invalid 		& Valid 		\\ \hline
		False Negative (FN) & Valid 		& Invalid 		\\ \hline
	\end{tabular}
\end{table}
\subsubsection{Training the Machine}\label{subsubsec:train_machine}
We consider the use of support vector machine (SVM) and adaptive boosting (AdaBoost) algorithms for \itapx~\cite{murphy2012machine}, as both these algorithms have been shown in the past to provide accurate performance in similar classification tasks~\cite{jaing2019machine}. We train our machine using synthetic data of $\Delta P_{\rsrp, i}$ and $\tau_{\text{q}, i-1}$ generated using different path loss models and introduce a higher penalty for false positives. We realize this by weighting false positives higher in our loss function minimization to target a $p_\text{TN} \geq 90\%$. Recall that false positives result in scenarios where the UE uses an invalid TA for CG-SDT, which causes increased interference with neighboring resource transmissions. This makes the use of CG-SDT a net negative to overall system functioning when compared to not using CG-SDT. However, a false negative leads only to a conservative self-deterrence, whose operation is equivalent to a regular transmission, i.e., a transmission that does not use the CG-SDT configuration. 

We randomize the movement of the UE so that its location, $\ell_i$, is drawn from a random variable $X$, where 
\begin{equation}\label{eq:UE_location_uniform}
    X \sim \mathcal{U}(0, d_{\max}),
\end{equation}
and $d_{\max}$ represents the maximum distance of the UE from the base station. This ensures that the generated $\ell_i$s are not correlated and that our machine is not trained for any set trajectory of motion. By randomizing the locations of the UE, the UE velocity, UE movement trajectory, and CG-SDT periodicity do not impact our emulation of UE positions during CG-SDT. However, the UE velocity does impact the RSRP measurement within the UE due to varying Doppler shift effects on UEs with different velocities. \textcolor{black}{We emulate various types of UEs, such as pedestrian UEs, slow moving vehicles, and UEs aboard high-speed automobiles, by using a range of different UE velocities. Recall that the computation of $\Delta P_{\rsrp, i}$ is up to the implementation of the UE according to the existing standard~\cite{ts_36214}. This gives us the ability to devise efficient standard-compliant RSRP computation techniques. To this end, we explore noncoherent combining techniques to improve the reliability of the measured $\Delta P_{\rsrp, i}$.}

We use the downlink reference signal transmitted periodically by the base station to determine RSRP. Using a large window for combining downlink reference signal REs assists in reducing the impact on non-idealities, such as Doppler effects and noise, on the measured RSRP. However, the larger the window used, the higher is the power consumption in the UE and the longer is the time-to-measure of RSRP. The length of the window required to achieve a satisfactory performance (i.e., for our trained machine to achieve our target $p_{\text{TN}}$ and $p_{\text{TP}}$) depends on the UE velocity and the operating SNR. We evaluated the precision of the measured RSRP for different UE velocities between stationary to $120$~km/hr and varying operating SNRs for UEs that are close to the base station to up to $5$~km away from the base station in urban macro and rural environments. We observed that using $4$ slots or subframes of the downlink reference signal that are each picked once every $10$~ms apart for combining provides a suitable trade-off between measurement accuracy and computation complexity or time-to-measure. Therefore, we use this setting throughout training and testing the performance of our method. We show in Section~\ref{sec:results} that this setting allows us to meet our target performance.  

\subsubsection{Tuning the Machine}
One of the goals of using ML for \itapx~is to ensure that our solution is applicable in a variety of path-loss environments irrespective of what was used during training. To verify this, we evaluate the robustness of our method by training our machine using one path loss model and testing its performance against a range of others. This also provides insight into our design of training the machine using synthetic data and deploying it in the real world. To better adapt to practical model deviation, we also scale the calculated $\Delta P_{\rsrp, i}$ before entering it into our classifier. Consider a machine that is trained using a path loss model with $k_{\text{train}}$ and implemented in an environment with a path loss exponent $k_{\text{system}}$. We scale the RSRP difference that is input to the machine as 
\begin{equation}\label{eq:rsrp_scaling}
	\Delta P_{\rsrp, i, \text{dB}} \leftarrow \Delta P_{\rsrp, i, \text{dB}} \frac{k_{\text{train}}}{k_{\text{system}}}. 
\end{equation}
This scaling ensures that $\Delta P_{\rsrp, i, \text{dB}} \propto k_{\text{train}}\log \left(\frac{d_i}{d_{i-1}}\right)$ as seen by the machine during training. We observed that this scaling operation provides a noticeable improvement in classification accuracy when operating in environments that were different from the ones used during training. However, this scaling operation requires the signaling of an estimate of $k_{\text{system}}$ to the UE by the base station that can be done either during the CG-SDT configuration and/or updated at any time during operation. \textcolor{black}{Note that this scaling only assists in countering the mismatch of path loss models. Even with an accurate path-loss model estimated by the base station and signaled to the UE, our designed method must still be robust in countering the RSRP measurement inaccuracies introduced due to fading effects and due to the impact of noise at low operating SNRs. } 

\subsubsection{Choice of \itapx}
We evaluate the performance of \itapx~in Section~\ref{sec:results} for both modes of operation, i.e., using the closed-loop closed-form expression-based \textit{threshold-based} method and the ML-based approach. We leave the choice of \itapx~mode to UE implementation depending on the performance accuracy and the implementation complexity trade-off that is desired. For example, a low complexity NB-IoT UE can choose to use threshold-based \itapx, whereas achieving superior classification performance can be prioritized for a 5G machine-type communication (MTC) device.

\section{TA Prediction}\label{sec:itaps}
TA validation, either using \itapx~shown in Section~\ref{sec:itapx} or the state-of-the-art method~\cite[Sec. 5.27]{tr_38321}, is efficient in allowing UEs to determine conditions when they can use an old TA for a new CG-SDT. This technique is suitable for UEs in relatively smaller cells. As cell size increases\footnote{An increase in cell size can be conceptualized either as an increase in $r_\cell$ or as using a carrier frequency in a higher frequency range (e.g., ``FR2'') and a subsequent increase in subcarrier spacing, with the same cell size.}, the probability of UEs moving beyond PM increases. In such cases, the UE must fall back on legacy random access techniques, with or without RA-SDT, to acquire a new TA from the base station. This limits the use of CG-SDT in such cases. To counter this, we propose an alternative method that ensures that UEs can always use CG-SDT. In lieu of validating TA, we propose the use of \itaps~to instead predict the TA at any given location of the UE so that it can use the predicted TA for CG-SDT. 

\textcolor{black}{TA prediction or estimation is by itself not new. Present-day 3GPP standards allow for TA estimation for radio access via non-terrestrial networks. However, they rely on the knowledge of location of the UE and the base station~\cite[Cl. 16.14]{tr_38300},~\cite{wang2021location}. Such methods add overhead in terms of signaling and/or cost at the UE. For example, location-based TA estimation that is currently used in non-terrestrial 3GPP networks requires the UE to determine its location using global navigation satellite system (or other positioning methods), along with receiving information about the location of the base station or the satellite. On the other hand, \itaps~requires no knowledge of where the UE or the base station is positioned and works solely based on the received signal strength of reference signals that are already transmitted by the base station for synchronization purposes.}

TA prediction using \itaps~in place of TA validation also eliminates the restriction of CG-SDT usage to movement within the PM. It allows system-wide implementation irrespective of the transmission parameters such as the OFDM sub-carrier spacing or environment characteristics like the cell-size. 
A 3GPP-standards-compliant system implementation of this technique is straightforward to achieve by letting the network set the CG-SDT-time alignment timer to $\infty$ and allow UEs to use \itaps~for TA determination. 

\textcolor{black}{Using \itaps, a UE begins by predicting a TA for any given UE location where the UE has a small amount of data to transmit using CG-SDT resources using current and historical RSRP values and a previous TA. We describe the prediction techniques in detail in Section~\ref{subsec:eq_itaps} and Section~\ref{subsec:ml_itaps}. The UE then uses the predicted TA to perform a CG-SDT. Upon successful reception of the CG-SDT packet, the base station provides the UE with the actual TA of the transmission. The UE then updates its currently held TA with the one received from the base station, which it then uses for subsequent TA prediction. The UE thereby always uses an actual TA from previous iterations for predicting a current TA. Therefore, any error in predicting a TA does not carry forward or compound over predictions, since the UE always receives an accurate TA value from the base station after every CG-SDT.}

\begin{figure}
    \centering
    \includegraphics[width=.8\linewidth]{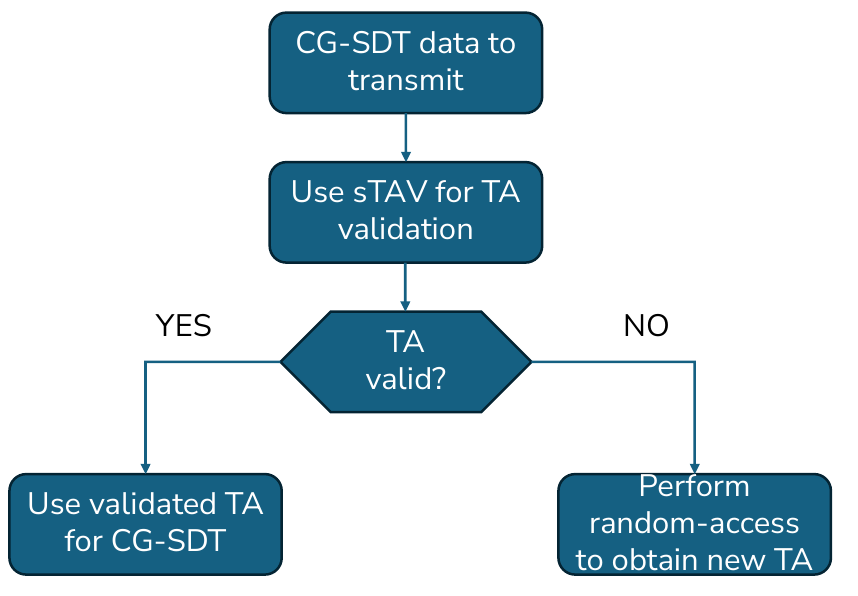}
    \caption{\textcolor{black}{Flow chart illustrating CG-SDT using sTAV.}}
    \label{fig:sTAV_flow}
\end{figure}

\begin{figure}
    \centering
    \includegraphics[width=0.35\linewidth]{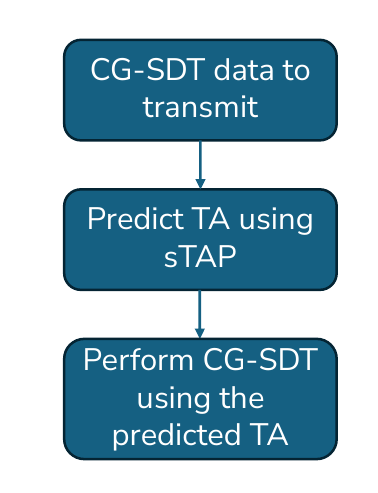}
    \caption{\textcolor{black}{Flow chart illustrating CG-SDT using sTAP.}}
    \label{fig:sTAP_flow}
\end{figure}

Using \itaps~reduces the average fallback rate for CG-SDT. \textcolor{black}{This can be visualized using the flow charts presented in Fig.~\ref{fig:sTAV_flow} and Fig.~\ref{fig:sTAP_flow}.} The average fallback rate can be defined as the probability of UEs falling back on legacy methods (e.g., moving to the connected mode after a random access procedure, or using RA-SDT) for TA acquisition. Falling back on legacy (i.e., non-CG-SDT) approaches is costly for a UE due to the increased power consumption associated with acquiring a new TA, e.g., with a random access procedure, when compared to a CG-SDT or PUR transmission~\cite{hoglund20203gpp}.

We categorize the UE fallback on to legacy methods into the following two types:
\begin{enumerate}
    \item Proactive fallback: In case of proactive fallback, the UE falls back to a legacy channel access procedure based on determining that the TA held by the UE is inaccurate. With the use of \itapx, the UE proactively falls back when the \itapx~indicates $\hat{\tau}_{\text{q},i} = \infty$.
    \item Reactive fallback: The UE falls back in a reactive manner when the UE uses an invalid TA, which leads to an unsuccessful CG-SDT. Upon learning of such an unsuccessful transmission (e.g., either by the base station indicating a failed detection or by the UE conducting repeated unsuccessful CG-SDT attempts), the UE falls back to legacy methods for channel access. 
\end{enumerate}
We denote the proactive and reactive fallback rates for \itapx~and \itaps~as $p_{\text{f,pro},\itapx}$, $p_{\text{f,re},\itapx}$, $p_{\text{f,pro},\itaps}$, and $p_{\text{f,re},\itaps}$, respectively. We then define the overall fallback rates for the \itapx~and \itaps~methods as
\begin{align}\label{eq:fb_itapx}
    p_{\text{f},\itapx} &= p_{\text{f,pro},\itapx} + p_{\text{f,re},\itapx} \\ \label{eq:fb_itaps}
    p_{\text{f},\itaps} &= p_{\text{f,pro},\itaps} + p_{\text{f,re},\itaps},
\end{align}
respectively. 

Next, we characterize the fallback rates based on the performance of \itapx~and \itaps. 
Using \itapx, the two events where the UE proactively falls back are when the UE has moved beyond the PM with \itapx~correctly indicating an invalid TA, and when the UE has moved within the PM with \itapx~incorrectly indicating an invalid TA. \textcolor{black}{This is shown in the ``{NO}'' path of Fig.~\ref{fig:sTAV_flow}.} Therefore, we can represent the proactive fallback rate of \itapx~as
\begin{equation} \label{eq:profb_itapx}
    p_{\text{f,pro},\itapx} = p_{\Delta d>\Delta d_\per} p_{\text{TN}} + (1-p_{\Delta d>\Delta d_\per})p_{\text{FN}},
\end{equation}
where $p_{\Delta d>\Delta d_\per}$ is the probability that the UE moves beyond the PM for using the previously held TA but remains within the same serving cell. The condition where the UE reactively falls back is when the UE incorrectly uses an invalid TA and learns about the TA validity after one or more unsuccessful CG-SDT attempts. Therefore,
\begin{equation}\label{eq:refb_itapx}
    p_{\text{f,re},\itapx} = p_{\Delta d>\Delta d_\per} p_{\text{FP}}.
\end{equation}
Finally, using \eqref{eq:FPR}, \eqref{eq:fb_itapx}, \eqref{eq:profb_itapx}, and~\eqref{eq:refb_itapx}, we determine
\begin{equation}\label{eq:fallback_validation}
    p_{\text{f},\itapx} = p_{\Delta d>\Delta d_\per} + (1-p_{\Delta d>\Delta d_\per})p_{\text{FN}}.
\end{equation}
On the other hand, using \itaps, we eliminate the cases of the UE proactively falling back to the legacy method since the UE always performs a CG-SDT using the predicted TA. \textcolor{black}{As seen in Fig.~\ref{fig:sTAP_flow}, UE always uses the predicted TA to perform CG-SDT without proactively falling back to perform a random-access procedure.} Therefore, the only fallback cases are when the UE inaccurately predicts the TA and reactively falls back. Therefore,
\begin{equation}\label{eq:fallback_prediction}
p_{\text{f},\itaps} = p_{\text{f,re},\itaps} = 1- \eta,
\end{equation}
where $\eta$ is the prediction accuracy of \itaps. Recall that the cyclic prefix provides a \textit{buffer} for using an inaccurate TA that is quantified in terms of the PM. Therefore, the TA prediction accuracy must only be equivalent to a UE movement within the PM. That is, if the predicted TA corresponds to an estimated UE-base station distance of $\tilde{d}_i$ and the actual UE-base station separation is $d_i$, the TA prediction is accurate if $\Delta d_i = |d_i - \tilde{d}_i| \leq \Delta d_\per$. Therefore,
\begin{equation}\label{eq:eta}
	\eta =  \frac{1}{2N_{\text{obs}}} \sum \limits_{i=1}^{N_\text{obs}} \left( 1 + \mathrm{sgn}(\Delta d_\per - \Delta d_i) \right),
\end{equation}
where $\mathrm{sgn}(x)$ is a modified signum function, which we define as
\begin{equation}
	\mathrm{sgn}(x) \triangleq 
	\begin{cases}
	+1, &\text{if } x = 0 \\
	\frac{|x|}{x}, &\text{otherwise } 
	\end{cases}
\end{equation}
and $N_{\text{obs}}$ is the number of observations over which the accuracy is computed. When $\Delta d_i > \Delta d_\per$, CG-SDT fails due to the use of an inaccurate transmission. This condition is equivalent to false positives in TA validation where the validator wrongly indicates an invalid TA to be valid. Therefore, similar to the TNR in \itapx, we target $\eta \geq 0.9$ for \itaps. 

\subsection{Equation-based \itaps}\label{subsec:eq_itaps}
\textcolor{black}{We use the threshold-based \itapx~presented in Section~\ref{subsec:threshold_itapx} as the baseline for what we refer to as \textit{equation-based} \itaps. Along the same lines as in Section~\ref{subsec:threshold_itapx},} we build a closed-form-expression-based \itaps~method using pre-defined and well-developed path-loss models. Estimating the TA after any UE movement can be drawn directly from the analysis in Section~\ref{sec:itapx}. Using~\eqref{eq:ta_quant} and~\eqref{eq:delRSRP_simple}, we compute the TA at the $i$th UE location as
\begin{equation}\label{eq:eq_itaps}
	\hat{\tau}_{\text{q},i} = \left \lfloor 10^{\frac{\Delta P_{\rsrp,i,\text{dB}}}{k}} \frac{2d_{i-1} }{c \tau_\step} \right \rfloor,
\end{equation}
where $\hat{\tau}_{\text{q},i}$ is the TA predicted by \itaps~for the $i$th CG-SDT (similar to the output TA in~\eqref{eq:itapx_output}).
\textcolor{black}{The value of $d_{i-1}$ used in~\eqref{eq:eq_itaps} is computed as}
\begin{equation}
    d_{i-1} = \frac{c\tau_{\text{step}}{\tau}_{\text{q},i-1}}{2},
\end{equation}
\textcolor{black}{where ${\tau}_{\text{q},i-1}$ is the actual TA value provided by the base station after a successful CG-SDT at the $(i-1)$th instance.}

This method introduces little computational complexity over the threshold-based \itapx~to directly estimate the TA. The signaling requirements associated with the use of this method are also the same as those for the threshold-based method, i.e., the path-loss exponent is conveyed to the UE by the base station either during CG-SDT configuration or any time during operation such as along with the CG-SDT response. However, due to the nature of the estimation that relies significantly on the accuracy of the path-loss model and the accurate computation of $\Delta P_{\rsrp,i,\text{dB}}$ within the UE, it suffers from the same drawbacks as the threshold-based \itapx. To address these challenges, we investigate the use of data-driven methods in the ML-aided \itaps.

\subsection{ML-aided \itaps}\label{subsec:ml_itaps}
\textcolor{black}{For the design of our ML-aided \itaps, we again draw from the ML-aided \itapx~as the baseline.}
\subsubsection{\textcolor{black}{Single-TA based ML-aided \itaps}}\label{subsec:ml_itaps_single}
\textcolor{black}{We begin by presenting \itaps~that runs using a single prior TA value for predicting the current TA. We formulate the problem as a supervised ML regression task where a regressor uses the previous TA, the previously determined RSRP, and the current RSRP measured by the UE to estimate the TA at the new UE location, i.e., }
\begin{equation}\label{eq:ML_stap}
\color{black}
	\hat{\tau}_{\text{q}, i} = \itaps(\tau_{\text{q}, i-1}, \Delta P_{\rsrp, i, \text{dB}} ).
\end{equation}   

We use the least squares boosting (L2Boost) algorithm to train the regressor. L2Boost consolidates multiple weak learners into a strong learner by applying weak learners to weighted versions of the data with higher weights allocated to samples suffering greater inaccuracies in previous rounds~\cite[Ch. 16]{murphy2012machine}. Additionally, it is also robust to over-fitting, which is particularly beneficial to our case of unrestricted training samples that we generate synthetically using established path-loss models. 

\subsubsection{\textcolor{black}{Multiple-TA based ML-aided \itaps}}
\textcolor{black}{We have thus far described our methods (both \itapx~and \itaps) with the use of a single previous TA and RSRP measurement to estimate or validate a current TA value for CG-SDT. This approach is suitable for C-IoT application scenarios where CG-SDT data is highly infrequent. As a result, UE positions between subsequent CG-SDT attempts can be modeled to be uncorrelated. However, for scenarios where the periodicity of CG-SDT data is low with more recurrent traffic, e.g., asset tracking, inventory updating, or traffic platooning, we explore the use of continuously tracking TA to possibly improve TA prediction accuracy.} 

\textcolor{black}{We use a long short-term memory (LSTM) model~\cite[Ch. 10]{zhang2023dive},~\cite[Ch. 15.2.7.2]{murphy2022probabilistic} to  predict TA values for different types of UE movements. Such a multi-input regression uses historical raw data of the measured RSRP and prior TA indications provided by the base station to estimate the new TA. We can therefore rewrite~\eqref{eq:ML_stap} as}
\begin{align} \label{eq:multiTA_itaps}
\color{black}
\hat{\tau}_{\text{q}, i} = \itaps({\tau}_{\text{q}, i-k},  P_{\rsrp, i-k, \text{dB}} ,
\forall k \in [\![1 \dots K]\!]),
\end{align}
\textcolor{black}{where $K$ is the total number of historical values used, and ${\tau}_{\text{q}, i-k}$ represents the TA value received from the base station after CG-SDT at the $(i-k)$th instance. We list all the parameters used in our LSTM architecture in Table~\ref{table:lstm_settings}.} 

\begin{figure}
    \centering
    \subfigure[]{\includegraphics[width=0.2\textwidth]{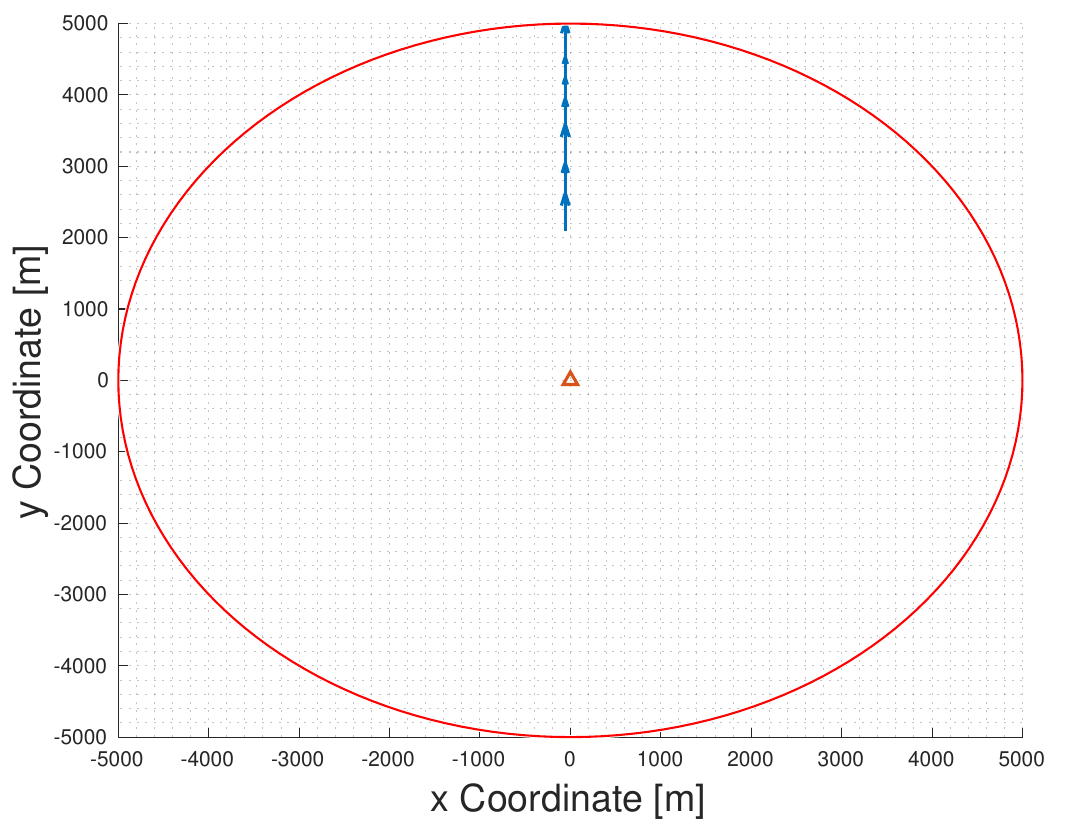}} 
    \subfigure[]{\includegraphics[width=0.2\textwidth]{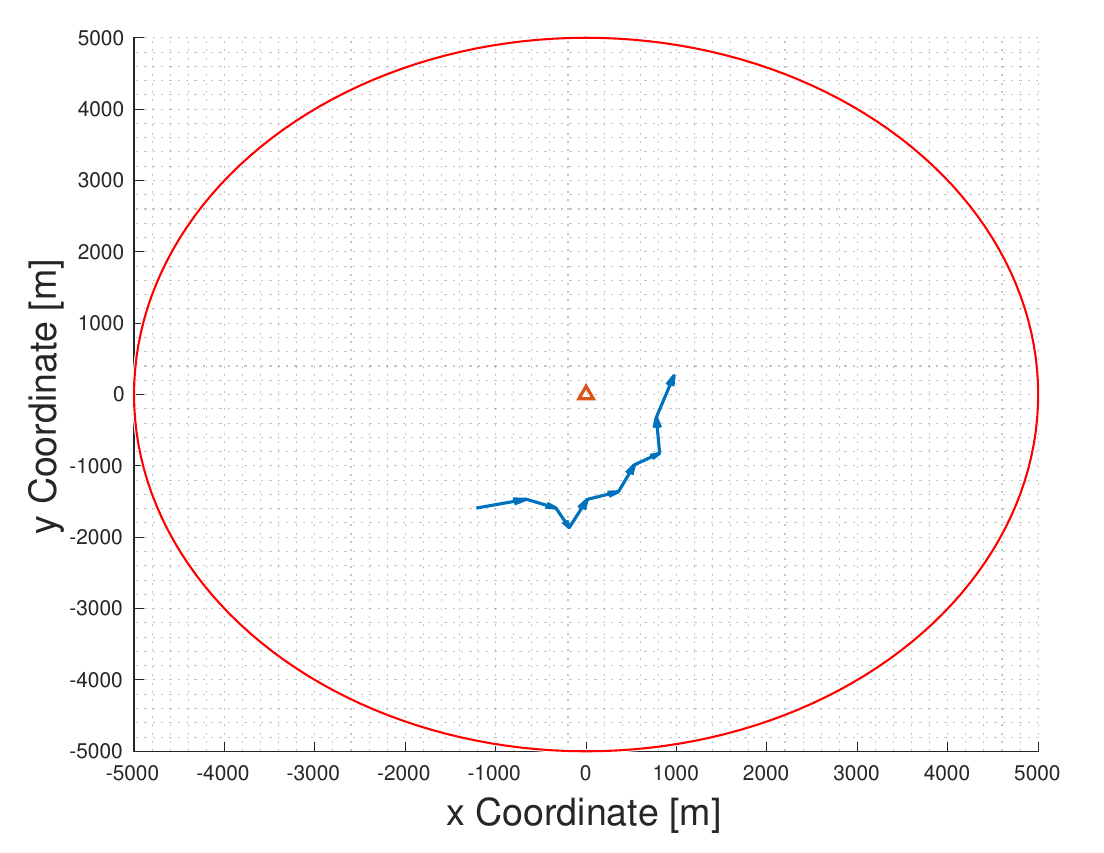}} 
    \subfigure[]{\includegraphics[width=0.2\textwidth]{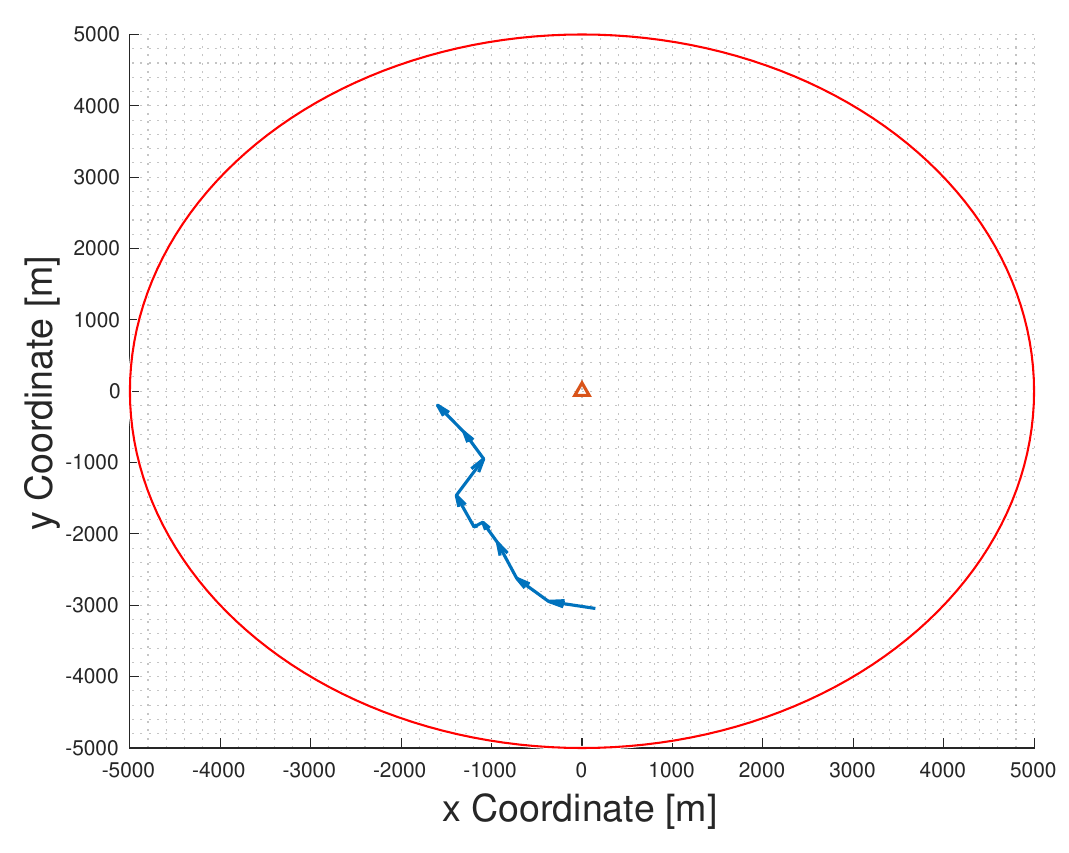}}
    \subfigure[]{\includegraphics[width=0.2\textwidth]{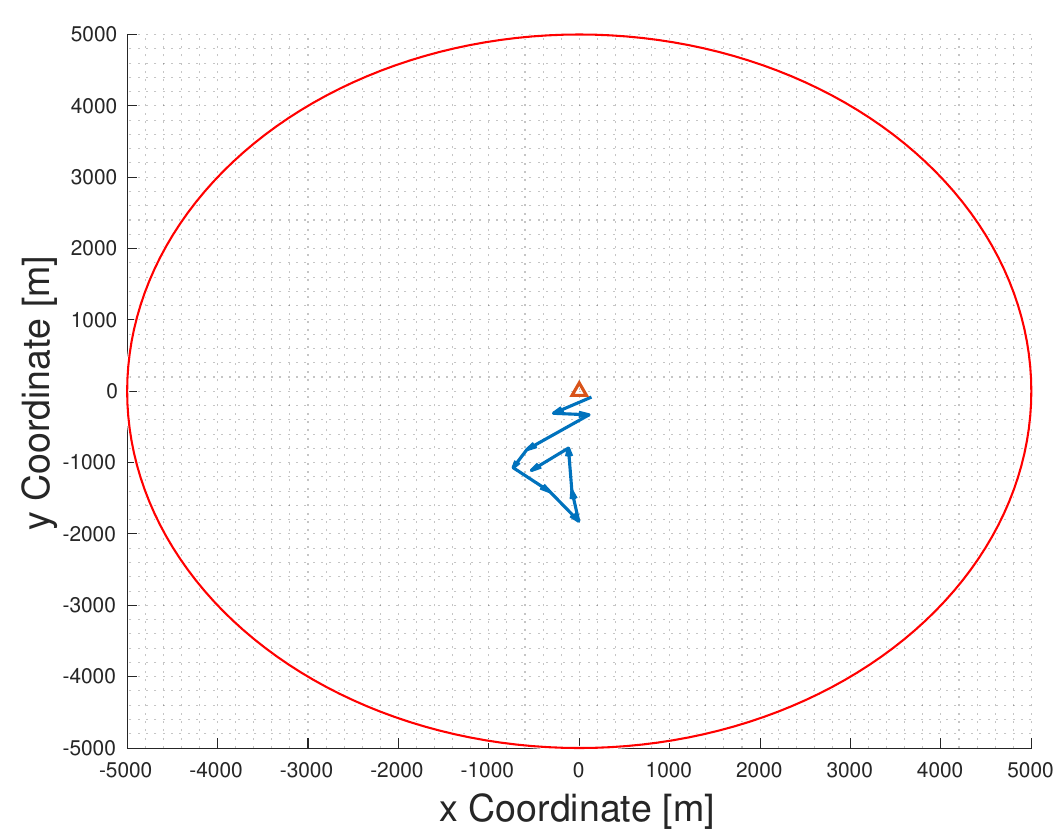}}
    \caption{Some non-exhaustive examples of UE movement with the added random shifts as the UE traverses along the following trajectories: (a) Manhattan or taxicab; (b) counter-clockwise; (c) clockwise; and (d) radial.}
    \label{fig:UE_movements}
\end{figure}

\textcolor{black}{Up until the previous subsection, we considered randomized UE locations in a two-dimensional space that is drawn from a uniformly distributed random variable as shown in~\eqref{eq:UE_location_uniform}. However, this is not suitable for a memory-based TA tracking approach, where using random uncorrelated UE locations may not provide any meaningful results. Such a method of distributing UE positions to be uniformly random is also not reflective of typical UE movements. Therefore, for our training and testing, we use four different UE trajectories with respect to the cell center: Manhattan or taxicab, counter-clockwise, clockwise, and radial paths. Furthermore, we apply  UE velocities, $v_\text{UE}$, based on various types of UE movements. We use four different configurations: (a) human walking and running, where $\overline{v}_\text{UE} = [5, 10]$~km/hr, (b) vehicles in residential areas, where $\overline{v}_\text{UE} = [20, 65]$~km/hr, (c) vehicles on motorways, where $\overline{v}_\text{UE} = [70, 120]$~km/hr, and (d) a combination of the above, where $\overline{v}_\text{UE} = [0, 120]$~km/hr.} \textcolor{black}{Additionally, to emulate realistic scenarios, we add a degree of randomness in the form of random-shifts to UE motion as the UE progresses along the set trajectories. Fig.~\ref{fig:UE_movements} shows examples of the four-types of trajectories we considered, i.e., Manhattan or taxicab in (a), counter-clockwise in (b), clockwise in (c), and radial paths in (d), and the impact on the applied random-shifts. As shown in the figures, the random-shifts often times introduce UE movements away from an ideal trajectory, which makes the path traversed by the UE more emblematic of real-world motion. }    

\subsection{{Choice of \itaps}}
\textcolor{black}{Selecting the right form of \itaps~can be based on the desired performance, its trade-off with UE complexity, and the type of traffic supported by the device. For example, similar to the choice decision of \itapx~in Section~\ref{sec:itapx}, lower complexity NB-IoT or \textit{RedCap} UEs~\cite{Veedu2022toward} may opt for the \textit{equation-based} \itaps~while 5G MTC devices may adopt the \textit{ML-aided} \itaps. Additionally, depending on the traffic type used by the UE, the network may control the number of historical TA samples to be used by the device for TA prediction. Using a single-TA \itaps~may be suitable when data traffic is sparse, whereas a more frequent and low periodicity CG-SDTs may use the multiple-TA based \itaps~for more accurate TA prediction using a multitude of previous actual TA values. We present these results and more in the following Section~\ref{sec:results}. }

\section{Results}\label{sec:results}
We present simulation results of the performance of our proposed \itapx~and \itaps~techniques to evaluate the accuracy of our validation and prediction, respectively. 

\subsection{Simulation Settings}
\begin{table}[t]
	\centering
	\caption{Fixed Simulation Settings}\label{table:sim_settings}
	\begin{tabular}{|l|l|}
		\hline
		Parameter  		 				& Value \\ \hline \hline
		Carrier Frequency 				& $0.9$~GHz \\ \hline
		System Bandwidth 				& $1.4$~MHz \\ \hline
		Transmit Power  				& $46$~dBm \\ \hline
		Receiver Noise Figure  			& $9$~dB \\ \hline
		Thermal Noise Floor 			& $-174$~dBm/Hz \\ \hline
		No. of Transmitter Antennas		& $1$ \\ \hline
		No. of Receiver Antennas 		& $1$ \\ \hline
	\end{tabular}
\end{table}
We perform link-level simulations by emulating the base station and UE transceivers in their entirety at the physical layer. We list the fixed settings of our simulations in Table~\ref{table:sim_settings}. Note that since only a downlink reference signal is of interest, we consider a unidirectional transmission link with the base station and the UE as the transmitter and receiver, respectively. 

We use the system model described in Section~\ref{sec:system_model}. We use a variable $r_\text{cell}$ between a typical urban cell radius of $1.5$~km and a larger macro cell with a radius of up to $5$~km. We use three different path loss models for large-scale fading: the urban micro (UMi), urban macro (UMa), and rural macro (RMa), which we obtain from the 5G channel models presented in TR 38.901~\cite{tr_38901}. To avoid repetitiveness, we do not regurgitate the path loss expressions, which can be found in~\cite[Sec. 7.4]{tr_38901}. We use the path-loss exponent of the testing model as an input to \itapx~and \itaps~when testing the robustness of our methods to perform under unseen conditions. Note that in a practical implementation, this corresponds to a \textit{best-guess} estimate of the path-loss exponent made by the base station and signaled dynamically to the UE.

We also use the tapped delay line (TDL) channel models to emulate the small-scale fading effects~\cite[Sec. 7.7]{tr_38901}. Since a mobile UE experiences different types of environments depending on its position in the cell, we use a varying line-of-sight probability computed as in~\cite[Sec. 7.4.2]{tr_38901}. Based on this, we pick the TDL model between TDL-A to TDL-E accordingly. Further, we simulate different types of communication environments using three different average delay spreads, the short, normal, and long spreads from~\cite[Sec. 7.7.3]{tr_38901}, for scaling the delays of the power delay profiles of our considered channel models. Additionally, to capture a wide range of application scenarios from stationary sensors to use in high-speed automobiles, we introduce a random UE speed of up to $120$~km/hr, which introduces Doppler effects that impact the measurement of $P_\rsrp$ as described in Section~\ref{subsubsec:train_machine}. \textcolor{black}{We further considered other radio link quality metrics such as reference signal received quality (RSRQ) and received signal strength indication (RSSI). However, none of the results presented in the following use these metrics since: (a) we did not observe any noticeable improvement in reducing the average fallback rate using RSRQ and/or RSSI when used in addition with RSRP; and (b) as we show in the following results, we are  able to achieve our target performance results with the use of only RSRP. }

\subsection{\itapx}\label{sec:results_validation}

We use TPR and TNR, i.e., sensitivity and specificity, respectively, of \itapx~as its key performance indicators. 
An ideal classifier attempts to maximize both the contradictory metrics of sensitivity and specificity. However, for a practical design, it is critical to consider the importance of each of these to develop strategies to sacrifice one to improve the other. A high specificity is critical in ensuring that the FPR is low and as a result, fewer instances of an invalid TA are used for CG-SDT. This reduces the inter-UE interference caused due to the use of an invalid TA. On the other hand, a higher FNR is less costly as it only causes redundant fallback but not interference with UEs using adjacent resources. Therefore, we target a specificity of $p_{\text{TN}} > 90\%$, which is consistent with targets proposed in prior arts~\cite{R4_1905499,R4_1905500}, and a sensitivity of $p_{\text{TP}} > 50\%$ such that CG-SDT is used more often than not under valid TA conditions. To this end, we apply a mis-classification cost for false positives that is five times higher than that for false negatives during training.

\subsubsection{Threshold-based \itapx}
\begin{figure*}[t]	
	\begin{center}
        \subfigure[]{\includegraphics[clip,width=0.75\columnwidth]{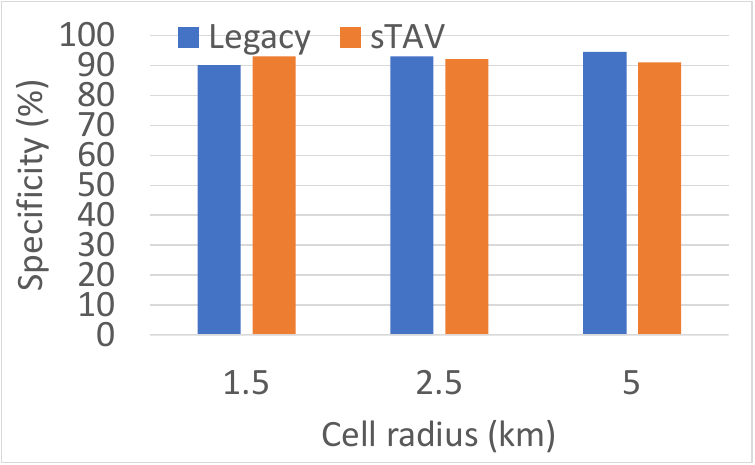}}
		\subfigure[]{\includegraphics[clip,width=0.75\columnwidth]{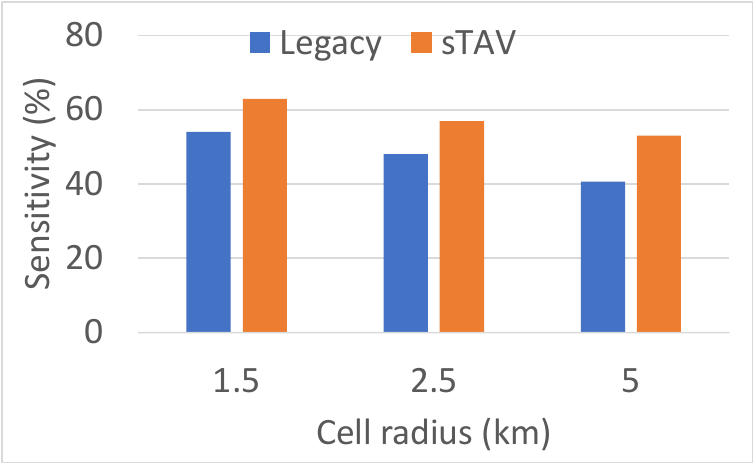}}
		\caption{The (a) specificity and (b) sensitivity of legacy and \itapx~methods for TA validation.}
		\label{fig:itapx_thresh}
	\end{center}
\end{figure*}
We present our first result of specificity and sensitivity of the threshold-based \itapx~and legacy TA validation in Fig.~\ref{fig:itapx_thresh}. We notice that while \itapx~matches the specificity of legacy methods, the use of \textit{smarter} error margins allows the sensitivity to be increased over the legacy validation method. For example, with an $r_\cell = 5$~km, legacy TA validation performs so poorly that around $6$ out of $10$ CG-SDT opportunities are wasted due to inaccurate validation results. On the other hand, the threshold-based \itapx~ exceeds the target TPR of $50\%$ in all evaluation conditions.

\subsubsection{ML-aided \itapx}
\begin{figure*}[t]	
	\begin{center}
		\subfigure[]{\includegraphics[clip,width=0.75\columnwidth]{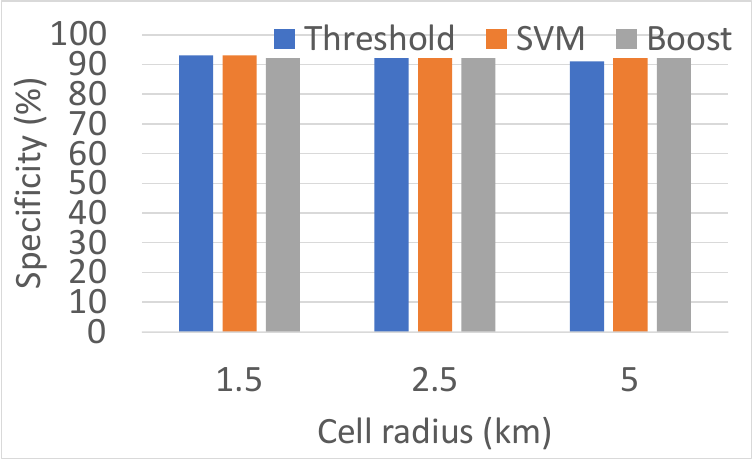}}
		\subfigure[]{\includegraphics[clip,width=0.75\columnwidth]{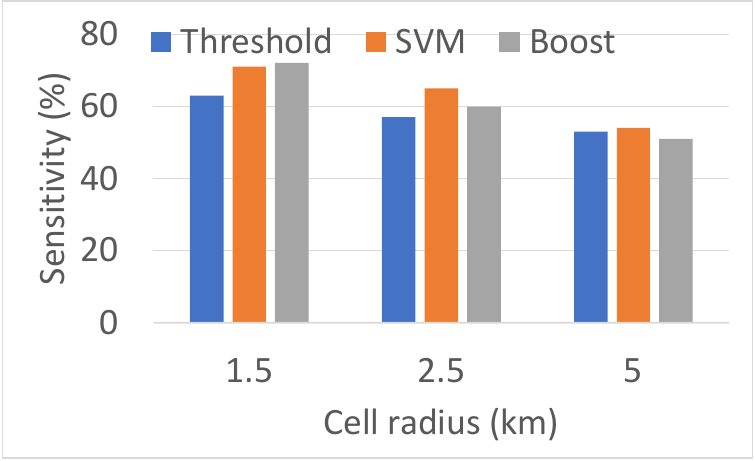}}
		\caption{The (a) specificity and (b) sensitivity of \itapx~methods for TA validation.}
		\label{fig:itapx_class}
	\end{center}
\end{figure*}
It is clear from Fig.~\ref{fig:itapx_thresh} that a threshold-based \itapx~provides the target specificity. However, our evaluation results of the ML-aided \itapx~shows the improvement in sensitivity achievable using our ML-aided approach. \textcolor{black}{We use SVM and AdaBoost with a false positive cost of $5$ and   $6$, respectively, to ensure that we match the target specificity of $90\%$, as explained in Section~\ref{sec:itapx}. The remaining model parameters and number of training and testing samples used for SVM and AdaBoost are summarized in Table~\ref{tab:table_number_samples}.} Fig.~\ref{fig:itapx_class} shows that ML-aided \itapx~using both SVM and AdaBoost match the specificity of the threshold-based \itapx~and provides a noticeable improvement in sensitivity. However, under large macro cells at $r_\cell = 5$~km, the AdaBoost based $\itapx$~falls behind the threshold- and the SVM-based $\itapx$ marginally, while still providing higher than a $50\%$ PUR usage rate.

\begin{figure*}[t]	
	\begin{center}
		\subfigure[]{\includegraphics[clip,width=0.75\columnwidth]{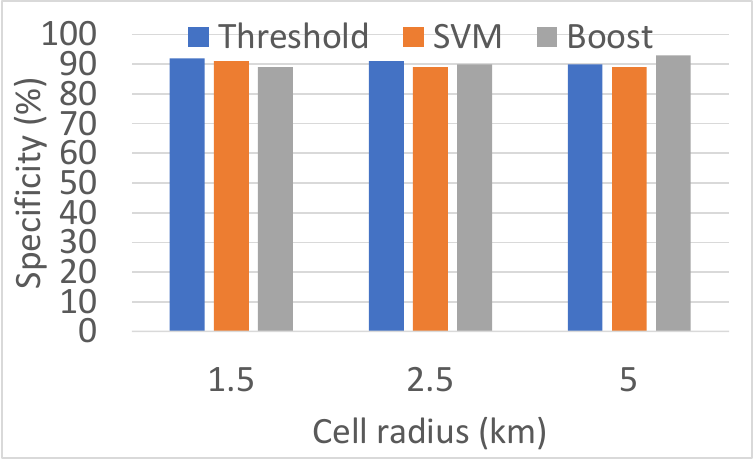}}
		\subfigure[]{\includegraphics[clip,width=0.75\columnwidth]{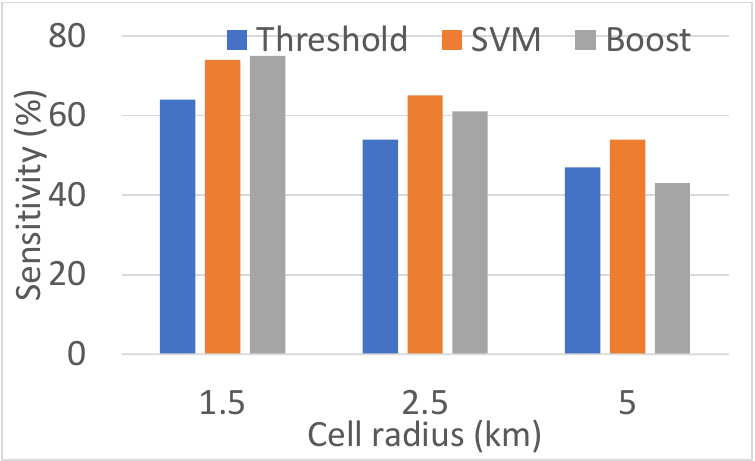}}
		\caption{The (a) specificity and (b) sensitivity of \itapx~methods for TA validation with cross path-loss model training and testing.}
		\label{fig:itapx_robust}
	\end{center}
\end{figure*}

Fig.~\ref{fig:itapx_robust} presents the same comparison as in Fig.~\ref{fig:itapx_class} for a more realistic evaluation setting, where the machines for the ML-aided $\itapx$~are trained and tested under different channel models. This captures practical conditions, where real-world channels need not necessarily fit any of the channel models that may be used during training. As an example, we used the urban macro path-loss model for training, and the urban micro path-loss model for the sub-2.5 km cell radius and the rural macro path-loss model for the 5 km cell radius case when testing. The positive results in Fig.~\ref{fig:itapx_class}, i.e., a $>90\%$ specificity and $>50\%$ sensitivity (except for the AdaBoost based $\itapx$) demonstrate that our proposed method can be used in an offsite training based deployment. An offsite training-based deployment consists of a setting where the machines are trained offline using synthetic computer-generated data, for example, using a best emblematic channel model of the potential environment, and deployed in the real world to function using practically extracted measurements within a UE. Together with eliminating the restriction on the number of training samples, such a training regimen also minimizes the \textit{startup} time that may be required by a \textit{learn-as-you-go} approach that incurs inaccurate performance initially until acquiring sufficient data to improve its performance.

\textcolor{black}{It is clear from the above results in Fig.~\ref{fig:itapx_thresh}(a), Fig.~\ref{fig:itapx_class}(a), and Fig.~\ref{fig:itapx_robust}(a) that our proposed \itapx~methods achieve our target performance goal of $90\%$ specificity. They also show that our results match that of the state-of-the-art techniques in terms of specificity. However, the benefit of our proposed methods over the state-of-the-art in seen in the sensitivity results in Fig.~\ref{fig:itapx_class}(b), Fig.~\ref{fig:itapx_thresh}(b), and Fig.~\ref{fig:itapx_robust}(b). Our~\itapx~methods outperform the legacy technique (marked as \textit{legacy} in Fig.~\ref{fig:itapx_class}), indicating that UEs using~\itapx~fallback less often to a random-access procedure for channel access. }

\subsection{\itaps}\label{subsec:singleTASTAP}

\begin{figure}[t]
	\centering
	\includegraphics[width=6cm]{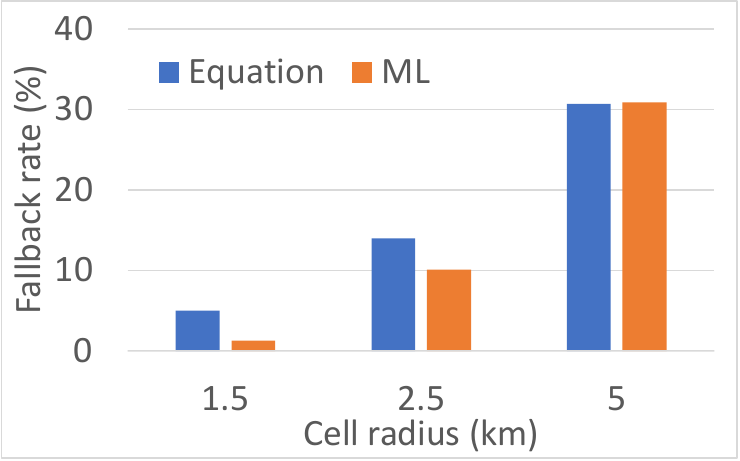} 
	\caption{The fallback rate for an equation-based $\itaps$~method and a single-TA based ML-aided $\itaps$~method with machines that were trained and tested on the same path-loss model.}
	\label{fig:itaps_robust_same}
\end{figure}

\begin{figure}[t]
	\centering
	\includegraphics[width=6cm]{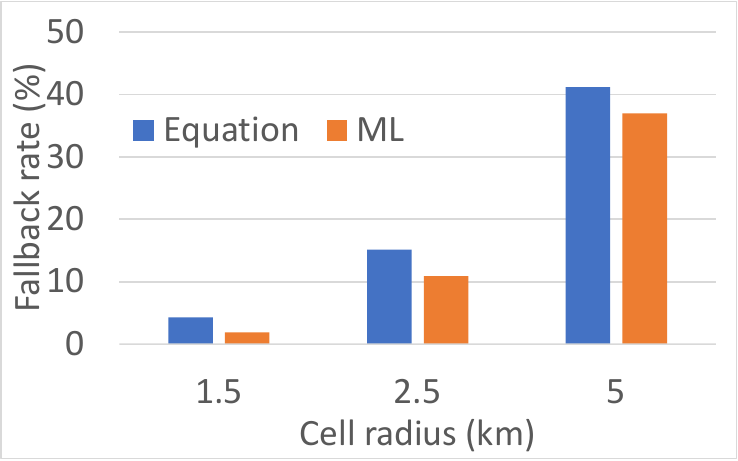} 
	\caption{The fallback rate for an equation-based $\itaps$~method and a single-TA based ML-aided $\itaps$~method with machines that were trained and tested on different path-loss models.}
	\label{fig:itaps_robust_diff}
\end{figure}

As explained in Section~\ref{sec:itaps}, \itaps~provides the ability for UEs to proactively estimate the TA value for CG-SDT so that it can be used at all required instances. We quantify the performance of our TA prediction solution using the fallback rate as defined in~\eqref{eq:fallback_prediction}. Fig.~\ref{fig:itaps_robust_same} and Fig.~\ref{fig:itaps_robust_diff} show the fallback rates for our proposed $\itaps$~solution with a uni-model and a cross-model training and testing approaches, respectively. We followed the same cross-model training and testing strategy as described in Section~\ref{sec:results_validation}. \textcolor{black}{The parameters for the L2Boost model and results are provided in Table~\ref{tab:table_number_samples}.} Our first takeaway from the results in Figs.~\ref{fig:itaps_robust_same} and~\ref{fig:itaps_robust_diff} is that the performance of our solution is robust across channel conditions drawn from path-loss models unseen during training. Second, using ML-aided \itaps~achieves fallback rates that are the same as or significantly lower than the fallback rates of the \textit{equation}-based \itaps. Third, the fallback rate from using our method is negligible when operating under typical urban cell sizes with a radius under a mile, especially with the ML-aided $\itaps$. Finally, we notice that the fallback rates increase with the radius of the cell with large cells encountering more instances where UEs fall back to random access procedure-based uplink transmissions for small data packets. 

\begin{table}[t]
	\centering
	\caption{Fallback rates for ML-aided \itapx~and single-TA based ML-aided \itaps}\label{table:fallback_rate}
	\begin{tabular}{|l|l|l|l|l|}
		\hline
		$d_{\text{cell}}$ (km)  & $p_{\text{FP}, \itapx}$   & $p_{\Delta d > \Delta d_\per}$ & $p_{\text{f},\itapx}$ & $p_{\text{f}, \itaps}$ 	\\ \hline \hline
		$1.5$ & $29\%$ & $25\%$ & $47\%$  & $1.9\%$ \\ \hline
		$1.5$ & $29\%$ & $50\%$ & $65\%$  & $1.9\%$ \\ \hline
		$1.5$ & $29\%$ & $75\%$ & $82\%$  & $1.9\%$ \\ \hline
		$2.5$ & $35\%$ & $25\%$ & $51\%$ & $11\%$ \\ \hline
		$2.5$ & $35\%$ & $50\%$ & $68\%$ & $11\%$ \\ \hline
		$2.5$ & $35\%$ & $75\%$ & $84\%$ & $11\%$ \\ \hline
		$5$ & $46\%$ & $25\%$ & $60\%$ & $37\%$ \\ \hline
		$5$ & $46\%$ & $50\%$ & $73\%$ & $37\%$ \\ \hline
		$5$ & $46\%$ & $75\%$ & $87\%$ & $37\%$ \\ \hline
	\end{tabular}
\end{table}
To provide context for the fallback rate numbers, we contrast our $\itaps$ values to the fallback probability observed with using a TA validation approach. The best performing TA validation method is the SVM-based $\itapx$~as seen in Section~\ref{sec:results_validation}. We use the FPR of SVM-based $\itapx$ in~\eqref{eq:fallback_validation} to compute the fallback probability of the validation method. Observe from~\eqref{eq:fallback_validation} that the fallback rate for TA validation further depends on $p_{\Delta d > \Delta d_\per}$. Greater the value of $p_{\Delta d > \Delta d_\per}$ for a UE, higher is the fallback rate, since TA validation attempts to invalidate and prohibit UEs from using CG-SDT when $\Delta d > \Delta d_\per$. On the other hand, our $\itaps$~solution proactively predicts the TA such that a UE can always use CG-SDT. 

We present the average fallback rates in Table~\ref{table:fallback_rate}. \textcolor{black}{We tabulate the results for three different values of $p_{\Delta d > \Delta d_\per} = \{25\%, 50\%, 75\%\}$ to account for different possible UE movements and CG-SDT intervals. For instance, a highly mobile UE with a large interval between successive CG-SDT occasions may have a higher $p_{\Delta d > \Delta d_\per}$ than a UE that moves slower with more frequent CG-SDT transmissions.} We observe that using \itaps~significantly reduces the fallback rates and improves the usability of CG-SDT. While we target an ideal zero fallback rate, which our ML-aided $\itaps$ comes close to in relatively small cells, any reduction in the fallback rate over the TA validation approach improves the applicability of CG-SDT. The worst-case fallback probabilities of \itaps, which is at $r_\cell = 5$~km, still cuts the fallback rates of $\itapx$~by more than half in highly mobile UEs. 

\begin{figure}[t]
	\centering
	\includegraphics[width=8.5cm]{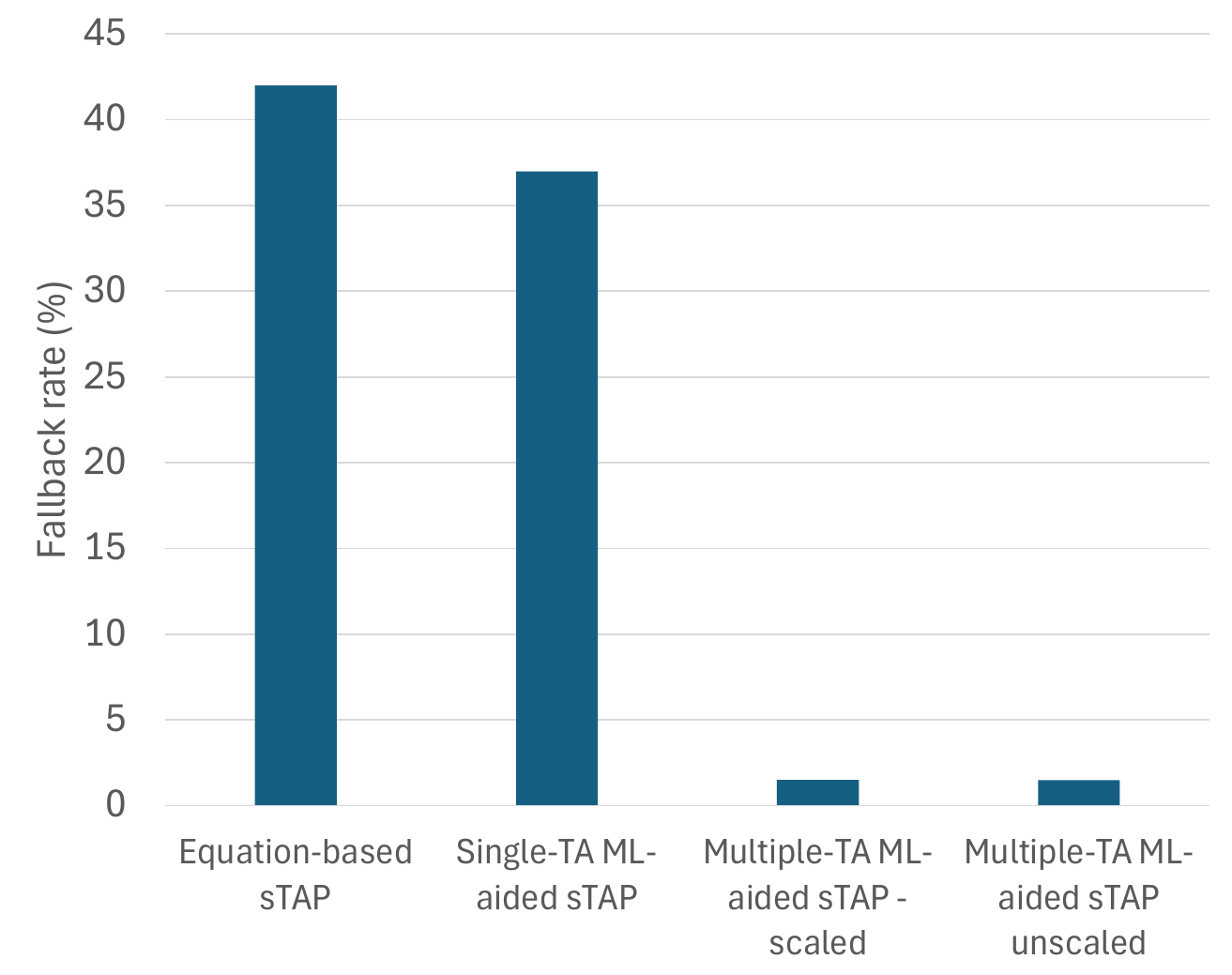} 
	\caption{The fallback rate for different variations of  $\itaps$~with machines that were trained and tested on different path-loss models for $r_\cell = 5$~km.}
	\label{fig:fallback_results}
\end{figure}

\subsection{Multiple-TA based ML-aided \itaps}\label{subsec:multiTA_prediction}
As our next result, we present the average fallback rate for \itaps~using more than one historical TA values for predicting the current TA. We focus on the case of $r_\text{cell}=5$~km, since we notice from the results presented in Section~\ref{subsec:singleTASTAP} that the performance of \itaps~in large-cell scenarios is the bottleneck for achieving low fallback rates. To this end, we use the LSTM model as described in Section~\ref{subsec:ml_itaps}. {\color{black}We present the corresponding LSTM settings and number of training and testing samples that we used for our simulations in Table~\ref{table:lstm_settings} and Table~\ref{tab:table_number_samples}, respectively.}

\begin{table*}
    \centering
    \color{black}
    \caption{Model parameters and the number of training and testing samples used.}
    \begin{tabular}{c|c|c|c}
        Model & Number of training samples & Number of testing samples & Model parameters \\ \hline \hline
         SVM (Fig.~\ref{fig:itapx_class} and Fig.~\ref{fig:itapx_robust})& $25,000$ & $10,000$ & radial basis function kernel \\ \hline
         AdaBoost (Fig.~\ref{fig:itapx_class} and Fig.~\ref{fig:itapx_robust})& $25,000$ & $10,000$ & decision trees, $150$ learning cycles \\ \hline
         L2Boost (Fig.~\ref{fig:itaps_robust_same} and Fig.~\ref{fig:itaps_robust_diff})& $25,000$ & $10,000$ & decision trees, $100$ learning cycles \\ \hline
         LSTM (Fig.~\ref{fig:fallback_results})& $1,565,850$ & $791,000$ & see Table~\ref{table:lstm_settings} \\ \hline
         LSTM (Fig.~\ref{fig:munich_fallback})& $242,978$ & $122,237$ &  see Table~\ref{table:lstm_settings}  \\ \hline 
    \end{tabular}
    \label{tab:table_number_samples}
\end{table*}

\begin{table}[t]
\color{black}
	\centering
	\caption{LSTM Settings}\label{table:lstm_settings}
	\begin{tabular}{|c|c|}
		\hline
        Number of features/input size & $2$ \\ \hline
        Hidden dimension & $5$ \\ \hline
        Number of epochs & $10$ \\ \hline
        Learning rate    & $0.01$ \\ \hline
        Batch size       & $64$ \\ \hline
        Loss function    & L1 \\ \hline
        Number of layers & 1 \\ \hline
        $K$ & $9$ \\ \hline
	\end{tabular}
\end{table}

\textcolor{black}{Fig.~\ref{fig:fallback_results} shows the average fallback rates for multiple-TA based \itaps, which we generated using over $130,000$ testing samples, in comparison with the equation-based and single-TA based ML-aided \itaps. For the multiple-TA based \itaps, we present two results, one with the scaling that we applied using~\eqref{eq:rsrp_scaling} and the other without scaling. Regardless of the application of scaling, we notice that using multiple-TA based \itaps~results in less than $2\%$ average fallback rate when compared to other forms of \itaps.} \textcolor{black}{This can be attributed to a number of factors. First, although we add randomness to UE trajectories, we still consider predictable paths taken by the UE, instead of random UE locations considered in the single-TA based ML-aided \itaps. Second, the use of $K > 1$ historical TA and RSRP values provides insight into the specific type of UE movement and the possible TA at a next instance, when compared to using a single previous TA value. Third, a sequence-based model like LSTM is able to exploit the structured data in the form of predictable UE movements when compared to non-sequence models like AdaBoost or SVM. Furthermore, we considered scenarios where UE trajectories are not unreasonably long. This ensures that we do not run into the exploding or vanishing gradient problems. By virtue of all these factors, our results demonstrate that using multiple-TA based \itaps~provides a significant improvement in TA prediction accuracy for CG-SDT scenarios.} 

\textcolor{black}{The results we have presented thus far demonstrates the effectiveness of our proposed solutions across different channel environments. However, we obtain all of these results using theoretical channel models like UMa, UMi, and RMa propagation models. To evaluate how well our results translate to real-world-like channel environments, we explored the use of ray-tracing based channel generation to generate propagation conditions in a practical environment. Ray tracing enables the simulation of channel realizations that are both environment-specific and physically accurate, representing a particular scene and user positions. We used Nvidia's Sionna 0.19 that provides a 
ray tracer for radio propagation modeling~\cite{hoydis2023sionna}. We consider an urban area in Munich featuring a $12$-meter base station mounted atop the $98$-meter \textit{Frauenkirche} tower. UE locations are randomly selected within a $1,500$-meter radius of the base station such that the path loss does not exceed $140$~dB. The maximum number of interactions between a ray and scene objects is set to $5$. A trajectory is constructed by connecting these points so that successive CG-SDT attempts correspond to UE positions spaced $200$ to $400$ meters apart, which is suitable considering the physical dimensions of the used scene. To render the task of TA prediction more challenging in this urban environment, 
we reduce PM from $700$~m to $175$~m. This scenario is emblematic of using higher sub-carrier spacing of $60$~kHz, as opposed to $15$~kHz sub-carrier spacing that we have consider so far. A sub-carrier spacing of $60$~kHz results in a shorter OFDM symbol length and a consequent shortening of the cyclic prefix, which in turn results in a smaller PM of $175$~m.   }

\begin{figure}[t]
    \centering
    \includegraphics[width=6cm]{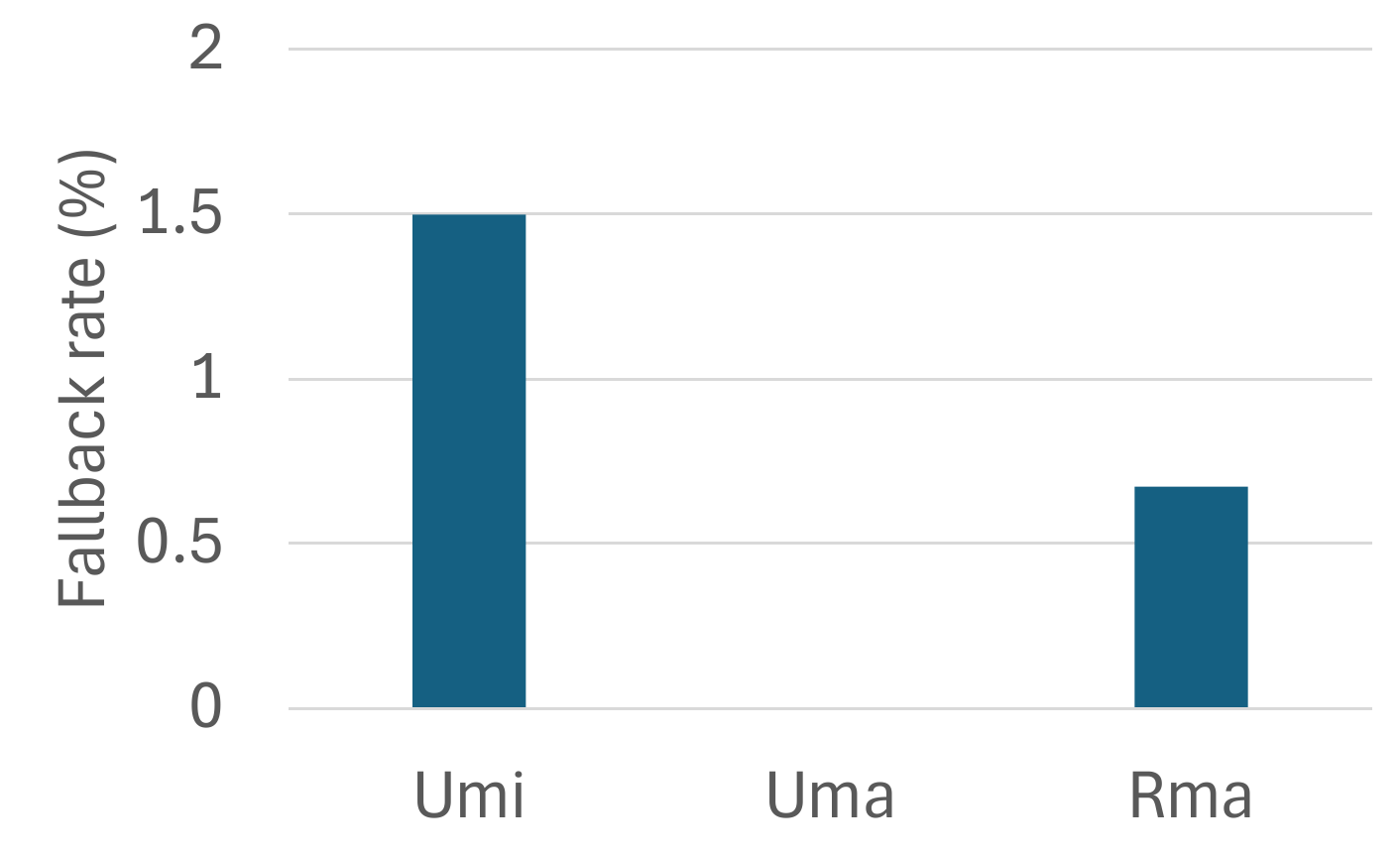}
    \caption{\textcolor{black}{Average fallback rates for UE movement in an urban area of Munich obtained with models trained using Umi, Uma, and Rma pathloss models.}}
    \label{fig:munich_fallback}
\end{figure}

\textcolor{black}{The average fallback rates that we obtained using such a setting is shown in Fig.~\ref{fig:munich_fallback}. We present results for the achieved fallback rates with models that are trained using UMi, UMa, and RMa propagation models, as used for our previous results. We observe that despite the fallback rates being non-zero for models that were trained with UMi and RMa channel models, they are significantly below what is achieved with the single-TA ML-aided \itaps~ and the equation-based \itaps~shown in Fig.~\ref{fig:fallback_results}. With the model that is trained with UMa propagation model, we observe no fallback to a random-access procedure, possibly owing to a closer match between the trained and tested channel conditions. The results in Fig.~\ref{fig:munich_fallback} clearly demonstrate the effectiveness of the scalability of our proposed solution to not just unseen channel conditions but also to ones that are emblematic of a real-world scenario.    }

\subsection{{Computational Complexity}}\label{subsec:computational_complexity}
\textcolor{black}{We conclude the section on performance results by analyzing the computational complexity of our proposed method and contrasting it against the state-of-the-art. We begin with~\itapx. The legacy TA validation method in~\cite[Sec. 5.27]{tr_38321} requires
\begin{equation}
    \gamma_{\text{legacy}} \approx 1 
\end{equation}
multiply and accumulate (MAC) operation. This assumes that RSRP values in dB scale are inherently computed by the UE for non-TA-validation purposes like channel state information reporting. Therefore, RSRP calculation can be excluded from $\gamma_{\text{legacy}}$ and the only required operation is to compute the difference in RSRP. }

\textcolor{black}{Our threshold-based \itapx~first involves computing $d_{i-1}$ to first determine the previous UE location. This is to check whether the UE is near the cell-edge or cell-center for using our improved thresholds as described in Algorithm~\ref{alg:itapx}. This requires about $\gamma_d = 2$ MACs. Next the UE performs between two to four comparisons after computing $\Delta P_{\rsrp, i, \text{dB}}$. This requires an additional MAC as with the legacy TA validation method. Since, comparisons do not typically require any MAC operations, our threshold-based \itapx~requires
\begin{equation}
    \gamma_{\text{\itapx, threshold}} = \gamma_d +\gamma_{\text{legacy}}  \approx 3
\end{equation}
MAC operations. }

\textcolor{black}{For ML-based \itapx, our proposed methods use AdaBoost and SVM. These are well-established ML models that have been used for similar classification tasks and are considered low-complexity implementations. Since SVM and AdaBoost perform very similarly in ML-aided \itapx, we select SVM for a more detailed complexity analysis. For each support vector, we need to perform two multiplications and one kernel evaluation. Hence, the complexity is $n_{\text s}(2+\gamma_{\text k})$ MACs, where $n_{\text s}$ is the number of support vectors and $\gamma_{\text k}$ is the number of MACs required for computing the kernel. As we use the radial basis function (RBF) kernel, its evaluation invokes $2d$ MACs for a norm and an inner-product computation and one exponential-function operation, where $d$ is the dimension of the feature vectors. The feature vector consists of the last valid TA and the change in RSRP, and therefore, $d=2$. The exponential-function evaluation can be implemented through a table look-up, polynomial, or other approximations. Denoting the corresponding number of MACs as $\gamma_{\exp}$, we may express the overall complexity as  
\begin{equation}
   \gamma_{\text{\itapx, ML-based}} = n_{\text s}(2+2d+\gamma_{\exp})=n_{\text s}(6+\gamma_{\exp}).
\end{equation}
The number of support vectors depends on the training data set, but might involve a few to several support vectors. Hence, we approximate the computational complexity as $\gamma_{\text{\itapx, ML-based}} < 100$.}

\textcolor{black}{Finally, since multiple-TA based ML-aided \itaps~is the winning variant of sTAP, we consider the complexity of LSTM. Accounting for the input gate, forget gate, output gate, cell gate, cell state update, and hidden state update operations~\cite[Ch. 10]{zhang2023dive}, for a sequence length of $K$,  we obtain the total MACs as 
\begin{equation}
    \gamma_{\itaps, \text{LSTM}} = 4Kd_h(d+ d_h) + 3Kd_h,
\end{equation}
where $d_h$ represents the hidden dimension and $d$ is the number of features. Substituting $K=9,~ d_h=5,~ d=2$ from Table~\ref{table:lstm_settings}, we get $\gamma_{\itaps, \text{LSTM}} = 1,359$. Complexity reduction methods, such as, replacing the LSTM with gated recurrent unit can further reduce $ \gamma_{\itaps, \text{LSTM}} = 3Kd_h(d+ d_h) + 3Kd_h= 1,080$.} 

\textcolor{black}{We observe that
\begin{equation}\nonumber
    \gamma_{\itaps, \text{LSTM}} \gg  \gamma_{\text{\itapx, ML-based}} \gg  \gamma_{\text{\itapx, threshold}} \approx \gamma_{\text{legacy}}.
\end{equation}
At the outset, this indicates that the ML operations introduce additional implementation complexity at the UE. While this is true when considering computations in isolation, we put these numbers into perspective by examining the overall power consumption by the UE for performing CG-SDT. Notice from Fig.~\ref{fig:fallback_results} that using \itaps~results in eliminating random-access fallbacks in a large majority of scenarios. This means that the UE avoids performing a random-access procedure, which involves at least two transmissions (i.e., random-access preamble and physical uplink shared channel transmission scheduled by random access response) and at least two receptions (i.e., random-access response and the contention resolution message). Radio transmissions and receptions consume power in the order of several mW~\cite{chandrika2021sync}, whereas performing $\gamma_{\itaps, \text{LSTM}} = 1,359$ MAC operations consumes a few nWs of power with a typical processor~\cite[Ch. 5]{fasthuber2013energy}. Therefore, the additional complexity introduced by LSTM introduces negligible overhead compared to the power savings achieved by nearly eliminating random-access fallbacks in CG-SDT scenarios.}

\section{Discussion} \label{sec:discussion}
Before concluding the paper, in this section, we briefly reflect on our proposed solutions by discussing their applicability in practical systems and providing guidelines for potential implementation designs.
\subsection{Standardization}
\textcolor{black}{3GPP has adopted the threshold- and RSRP-based TA validation in the LTE-M and NB-IoT specifications~\cite{ts_36331} as well as the 5G specifications~\cite{tr_38321, tr_38331}. Since the computation of RSRP is up to UE implementation, our threshold-based $\itapx$~method can be readily adopted into present-day C-IoT UEs. Further, the configuration of TA validation conditions for using PUR and CG-SDT is optional in the LTE-M, NB-IoT, and 5G specifications~\cite{ts_36331, tr_38331}. Therefore, network providers can choose to enable any of our $\itapx$~and/or $\itaps$~methods on compatible UEs without any impact to the present-day LTE-M, NB-IoT, or the 5G standards to counter the issue of model deficits for PUR transmissions and CG-SDT.}

\textcolor{black}{Aspects related to deployment and management of our proposed ML models are largely out of the scope of this paper. However, the ongoing 3GPP standardization work on life cycle management (LCM) for AI/ML models used for channel state information prediction, beam management, and positioning, are equally applicable to our designs~\cite{lin2023embracing, lin2023artifical}. Our proposed solutions are customizable at both the base station and UE-ends and therefore, the \textit{functionality} based LCM and the \textit{model identity} based LCM are both compatible with our design, along with the \textit{Level-Z} collaboration with model transfer~\cite{lin2023artifical}.}

\subsection{{Application of Multiple-TA based ML-aided \itaps}}
\textcolor{black}{Although exploring multiple-TA based \itaps~is motivated by scenarios with more frequent traffic, we note that we are not necessarily limited to lower periodic data. In those cases where the CG-SDT resource periodicity is low, UEs may choose to use actual TA values provided by the base station after successful CG-SDTs for predicting the current TA. However, UEs can estimate TA values at any point in between the CG-SDTs and as often as desired since SSBs are periodically broadcast by the base station. As a result, the periodicity of predicting TA values is not controlled by the CG-SDT resource periodicity, but instead by the periodicity of the downlink reference signals, i.e., the SSB measurement and timing configuration. Depending on the network implementation, this periodicity may be as low as $5$~ms~\cite{tr_38331}. }

\subsection{Implementation Consideration}
\textcolor{black}{Notice that the ML methods used for both \itapx~and \itaps~are well-established techniques that are designed to be robust to avoid common issues, such as, over-fitting. This choice is intentional. While prediction precision is important, TA estimation for CG-SDT is required to be accurate enough to the extent corresponding to about $702$~m for typical operating sub-carrier spacing of $15$~kHz, as described in Section~\ref{sec:system_model}. This allows us to focus on limiting computation complexity overheads without sacrificing prediction performance by using SVM and/or AdaBoost for ML-aided \itapx~and L2Boost and/or LSTM for ML-aided \itaps. We demonstrate this in Section~\ref{sec:results}, where are able to substantially improve TA validation sensitivity with \itapx~and nearly eliminate fallbacks using \itaps, all while introducing negligible additional power consumption.} 

The offline training-based deployment approach discussed in Section~\ref{sec:results} is one of several approaches of implementing our proposed solution. Training the machines can also be performed in the real world at the base station-end to ensure no increase in complexity at the C-IoT UEs. The RSRP and TA values can be collected by the base station from the several UEs in a given environment (e.g., a cell or a sector of the cell) to train a machine. The base station can exploit the reciprocal nature of frequency division duplex channels and use uplink reference signals such as the sounding reference signal to obtain the path-loss characteristics of the channel, similar to the architecture presented in~\cite{kim2021two}. 

\textcolor{black}{While PURs and CG-SDT configurations allow power savings at the UE, reserving dedicated radio resources to individual UEs may result in under utilization of network resources~\cite{khlass2021efficient}. \itaps~shows reduction in random-access fallback rates. This link-level reduction also translates to system level benefit by enabling better utilization of network resources, specifically, physical random-access channel resources. This allows for better resource utilization at the network side by offsetting some of the costs associated with reserved resource allocation. Network implementations can also further restrict the number of UEs that can be supported with CG-SDT operation by considering the trade-offs between reduced random-access resource usage with \itaps~and increased reservation of dedicated CG-SDT resources per UE. }

\section{Conclusion} \label{sec:conclusion}
We presented an autonomous timing synchronization solution for small data transmission in mobile C-IoT devices. Our UE-native solution includes two options for validating and/or predicting the TA at any given small data transmission instance. Our smart TA validator leverages historical path-loss measurements and trained ML models to intelligently predict whether a previously held TA is valid for SDT at a considered instance. We expanded on this design to present a smart TA predictor that exploits the same historical path-loss data to predict a TA at any SDT opportunity using ML-aided regression techniques. Our comprehensive simulation evaluation across a variety of channel conditions demonstrated the effectiveness of both of our design approaches in the real world. Measurement collection and prototype development to bring our proposed solutions to hardware implementation guide our future work.

\bibliographystyle{IEEEtran}		
\bibliography{References}{}


\end{document}